\documentclass[
    reprint,
    aps,
    prx,
    superscriptaddress,
    nofootinbib
]{revtex4-2}

\usepackage[utf8]{inputenc}
\usepackage[T1]{fontenc}
\usepackage{lmodern}
\usepackage[english]{babel}
\usepackage{graphicx}
\usepackage[hidelinks,colorlinks=true]{hyperref}
\usepackage{amssymb}
\usepackage{mathtools}
\usepackage{qcircuit}
\usepackage{bm}
\usepackage[caption=false]{subfig}
\usepackage{booktabs}
\usepackage{stmaryrd}

\DeclarePairedDelimiter\ket{\lvert}{\rangle}
\DeclarePairedDelimiterX\braket[2]{\langle}{\rangle}{#1 \delimsize\vert #2}

\newcommand{\im}{\mathrm{i}}

\newcommand{\calO}{\mathcal{O}}

\DeclareMathOperator*{\argmax}{arg\,max}

\def\quantinuumLondon{Quantinuum, Partnership House, Carlisle Place, London SW1P 1BX, United Kingdom}
\def\quantinuumTokyo{Quantinuum K.K., Otemachi Financial City Grand Cube 3F, 1-9-2 Otemachi, Chiyoda-ku, Tokyo, Japan}
\def\riken{Interdisciplinary Theoretical and Mathematical Sciences Program (iTHEMS), RIKEN, Wako, Saitama 351-0198, Japan}
\def\quantinuumCambridge{Quantinuum, Terrington House, 13-15 Hills Road, Cambridge CB2 1NL, United Kingdom}

\usepackage{soul}
\usepackage[dvipsnames]{xcolor}

\begin{document}

\title{Demonstrating Bayesian Quantum Phase Estimation with Quantum Error Detection}

\author{Kentaro Yamamoto}
\email{kentaro.yamamoto@quantinuum.com}
\affiliation{\quantinuumTokyo}

\author{Samuel Duffield}
\affiliation{\quantinuumLondon}

\author{Yuta Kikuchi}
\affiliation{\quantinuumTokyo}
\affiliation{\riken}

\author{David Mu\~noz Ramo}
\affiliation{\quantinuumCambridge}

\date{\today}

\begin{abstract}
Quantum phase estimation (QPE) serves as a building block of many different quantum algorithms and finds important applications in computational chemistry problems. Despite the rapid development of quantum hardware, experimental demonstration of QPE for chemistry problems remains challenging due to its large circuit depth and the lack of quantum resources to protect the hardware from noise with fully fault-tolerant protocols. 
In  the  present  work,  we  take a  step  towards fault-tolerant quantum computing by demonstrating a QPE algorithm on a Quantinuum trapped-ion computer. We employ a Bayesian approach to QPE and introduce a routine for optimal parameter selection, which we combine with a $\llbracket n+2,n,2\rrbracket$ quantum error detection code carefully tailored to the hardware capabilities.
As a simple quantum chemistry example, we take a hydrogen molecule represented by a two-qubit  Hamiltonian and estimate its  ground state energy using our QPE protocol.
In the experiment, we use the quantum circuits containing as many as 920 physical two-qubit gates to estimate the ground state energy within $6\times 10^{-3}$ hartree of the exact value.

\end{abstract}

\maketitle


\section{Introduction}

Classically hard computational chemistry problems are some of the most promising applications that the advance of quantum computing technologies will allow us to tackle~\cite{Aspuru-Guzik2005-tv,Cao2019-tn,McArdle2020-rs,Liu2022-ej}.
Chemical properties are generally calculated by solving the electronic structure problem for a given nuclear configuration~\cite{Helgaker2014-ce}.
Some systems (e.g., chemical reactions involving transition metal complexes \cite{Kim1992-qt,Umena2011-dr}) require many one-electron orbitals to represent their electronic states, making it challenging to accurately calculate their properties, even with state-of-the-art high-performance computers.
Harnessing the ability to efficiently simulate quantum systems, quantum computers will potentially facilitate those calculations to extend the scope of computational chemistry.

Quantum phase estimation is a heavily used subroutine in various quantum algorithms.
The QPE algorithm, provided a unitary operator $U$ with eigenvector $\ket{\phi}$ and associated eigenvalue $e^{i\phi}$, computes an estimate of the phase $\phi$.
The iterative scheme of QPE, originally proposed by Kitaev~\cite{kitaev1995quantum,kitaev2002classical}, runs a set of Hadamard-test-like circuits, each of which reads off the partial information about $\phi$. The obtained measurement outcomes are classically post-processed to estimate $\phi$ up to specified accuracy. Relative to the QPE based on the quantum Fourier transform (QFT)~\cite{Cleve:1997dh,Abrams1999,nielsen2002quantum}, the iterative scheme uses shallower circuits at the cost of more samples and hence, makes its execution amenable on noisy quantum hardware.
The iterative QPE algorithm has since been improved to reduce the required quantum resources~\cite{Knill2007, Higgins2007, Dobsicek2007, svore2013faster, Wiebe2016, OBrien2019, Somma2019, Sugisaki2021-me, Lin2022, Wan2022}. 
For example, the authors of~\cite{Wiebe2016} proposed an approach based on Bayesian inference to make the phase estimation protocol more efficient and robust against experimental errors.

Such algorithmic improvements, along with the remarkable development of quantum computing devices, have led to successful demonstrations of QPE algorithms on several different platforms.
Experimental realizations of QPE on a quantum chemistry system were initiated in~\cite{OMalley2016}, where the iterative QPE was applied to a two-qubit molecular Hamiltonian on a superconducting processor.
This was soon followed by an experiment using Bayesian QPE on a silicon photonic device~\cite{Paesani2017-dy}.
More recently, a different type of statistical phase estimation has been tested on a superconducting processor~\cite{Blunt:2023gqs}. In this algorithm, the trade-off between the number of samples and circuit depth is nicely controlled, which makes it more favourable to be run on early fault-tolerant hardware~\cite{Lin2022, Wan2022}.
However, experimental demonstration of QPE remains challenging due to the depth of the circuits.

In the present work, we demonstrate Bayesian QPE on a quantum charge-coupled device trapped-ion quantum computer to push forward such limitations~\cite{Wineland1997,Home2009,Monroe2013,Kaushal2020,Pino:2020mku}.
We calculate the ground-state energy of the minimal-basis molecular hydrogen system described by the spin Hamiltonian~\cite{Bravyi2017-xh},
\begin{equation}
\label{eq:hamiltonian_h2}
    H =
    h_{1}Z_{1}
    + h_{2}Z_{2}
    + h_{3}Y_{1}Y_{2}
    + h_{4}Z_{1}Z_{2}
    + h_{5}I,
\end{equation}
where $\{X_i,Y_i,Z_i\}$ are the Pauli-$X$, $Y$, $Z$ operators acting on $i$th qubit, $I$ is the identity operator, and $\{h_{j}\}$ are real coefficients. The spin Hamiltonian is obtained with the Jordan--Wigner transformation from its fermionic Hamiltonian~\cite{McArdle2020-rs}.\footnote{The coefficients $\{h_{j}\}$ are evaluated on classical computers by performing the restricted Hartree--Fock method with the STO-3G atomic basis set~\cite{Helgaker2014-ce}
followed by the two qubits tapered off using 
spin and particle number conservations~\cite{Bravyi2017-xh}.}

To enhance the performance of the Bayesian QPE protocol, we develop a subroutine for the selection of circuit parameters. Our routine analytically extracts the optimal parameters by minimising a natural choice of utility function. This improves on previous works based on heuristics with certain assumptions~\cite{Wiebe2016}.

We perform the QPE experiments by adopting two different techniques to alleviate hardware noise effects.
First, we employ the Bayesian QPE with a noise-aware updating rule. The hardware noise is incorporated into the likelihood function in the form of depolarizing channel to mitigate decoherence.
Second, 
in order to run the QPE with deep circuits, we encode four logical qubits with a $\llbracket 6,4,2\rrbracket$ quantum error detection code~\cite{Steane:1996va,Gottesman:1997qd,knill2004scheme,knill2004threshold}.
The code detects an arbitrary single-qubit error and discards the associated result to reduce the error rate at the cost of more circuit executions as well as small overhead in circuit depth and width.
In particular, 
the code is carefully tailored to various features of the quantum hardware, such as high-fidelity gate operations, all-to-all connectivity, and conditional exit operations~\cite{Pino:2020mku,Ryan-Anderson2022,Self2022}.
See~\cite{Linke2017,Takita2017,vuillot2017error,Roffe2018,Willsch2018,Harper2019,Urbanek2020,Zhang2022} for previous experimental studies with a $\llbracket4,2,2\rrbracket$ code.

The rest of the paper is organized as follows.
In Section~\ref{sec:BQPE}, we review the Bayesian QPE protocol, introduce our parameter selection rule and demonstrate its benefits numerically.
In Section~\ref{sec:QED}, the $\llbracket 6,4,2\rrbracket$ error detection code is introduced with a sketch of the encoded QPE circuit.
The experimental results from Quantinuum's H1 quantum computer are presented in Section~\ref{sec:experiment}.
We give conclusions and outlook in Section~\ref{sec:conclusion}.
Technical aspects of the Bayesian QPE, error detection code, experimental setup, and more data are found in Appendices~\ref{app:BPE}, \ref{app:QED}, and \ref{app:gate_error_mitigation}.

\section{Bayesian Quantum Phase Estimation}
\label{sec:BQPE}

In Bayesian QPE~\cite{Wiebe2016}, measurement outcomes are generated from the following QFT-free QPE circuit
\begin{align}
\label{circ:hadamard}
\begin{array}{c}
\Qcircuit @C=.5em @R=.7em {
    \lstick{\ket{+}}
    & \qw
    & \qw
    & \ctrl{1}
    & \gate{R_{Z}(\beta)}
    & \measureD{X}
    \\
    \lstick{\ket{\phi}}
    & {/} \qw
    & \qw
    & \gate{U^{k}}
    & \qw
    & \qw
}
\end{array}
\end{align}
parametrised by $k \in \mathbb{N}$ and $\beta \in [0, 2\pi)$, with probability
\begin{equation}
\label{eq:likelihood}
    p(m \mid \phi, k, \beta) 
    = \frac{1 + \cos(k\phi + \beta - m\pi)}{2},
\end{equation}
where $m \in \{0, 1\}$ is the measurement outcome in $X$-basis on the top qubit in \eqref{circ:hadamard} and $\phi \in [0, 2\pi)$ is the unknown angular phase we wish to determine. The equation \eqref{eq:likelihood} is termed the \textit{likelihood} function. $R_Z(\beta)=e^{-i\beta Z/2}$ is a rotation gate.
In our experiment, we consider an approximate eigenstate of $U$ as an input state $\ket{\phi}$ that has a sufficiently small approximation error.\footnote{When the input state is given by $\sum_ic_i\ket{\phi_i}$ with $U\ket{\phi_i}=e^{i\phi_i}\ket{\phi_i}$, we can use the likelihood of the form
\begin{align}
\label{eq:mix_likelihood}
    p(m \mid \{\phi_i\}, k, \beta) 
    = \frac{1 + \sum_i|c_i|^2\cos(k\phi_i + \beta - m\pi)}{2}.
\end{align}
Ignoring all $\{c_i\}_{i\ge1}$, where $c_0$ is the coefficient of the dominant eigenstate (e.g. ground state) $\ket{\phi_0}$, Eq.~\eqref{eq:mix_likelihood} reduces to Eq.~\eqref{eq:likelihood}.
The error in the estimated phase, $\phi_0$, due to the neglected contributions from excited states is bounded by $(1-|c_0|^2)/|c_0|^2$ in the case data is taken only from $k=1$. 
See e.g.~\cite{OBrien2019} for further discussion.
}


Taking a Bayesian approach, we instantiate a prior distribution $p(\phi)$, which can be uniform. After $R$ measurement outcomes, we probabilistically quantify the value of $\phi$ through the posterior distribution
\begin{equation}
\label{eq:posterior}
    p(\phi \mid m_{1:R}) \propto p(\phi) \prod_{r=1}^R p(m_r \mid \phi, k_r, \beta_r).
\end{equation}
The posterior can be updated iteratively $p(\phi \mid m_{1:R}) \propto p(\phi \mid m_{1:R-1}) p(m_R \mid \phi, k_R, \beta_R)$, until desired precision is obtained.

Bayesian posterior distributions are typically approximated numerically as in \cite{Wiebe2016}. However, in this setting, it was noted in~\cite{OBrien2019} that the posterior is analytically tractable in the form of a Fourier distribution 
\begin{equation}
\label{eq:fourier_posterior}
    p(\phi \mid m_{1:R}) = \frac{1}{2\pi} + \sum_{j=1}^{J_R} c_{R, j} \cos(j\phi) + s_{R, j} \sin(j\phi),
\end{equation}
where the number of coefficients $J_R$ grows at least quadratically in the number of experiments, as $J_R = J_{R-1} + k_R$. To mitigate memory requirements, the authors in~\cite{vandenBerg2021} described an adaptive approach where after a suitable number of shots, the posterior representation is converted to a wrapped Gaussian distribution which is fully specified by a mean and variance parameter and can also be updated analytically (using moment-matching) on receipt of further measurements. The conversion from Fourier to wrapped Gaussian distribution is approximate, however after a sufficient number of shots, the posterior will be peaked and well approximated by the wrapped Gaussian distribution. In this work, we adopt an equivalently adaptive approach only we favour a von Mises distribution over the wrapped Gaussian since it is the maximum entropy distribution for wrapped random variables (with a specified first circular moment) \cite{Jupp2009} and has a simple closed-form probability density function. A von Mises representation of the posterior is similarly fully specified by a mean and precision parameter and can also be updated analytically using moment-matching. Some example posterior distributions are visualised in Fig.~\ref{fig:posterior5}, Appendix~\ref{app:posteriors}.

Details of the tractable Bayesian updates can be found in Appendix~\ref{app:BPE_updates} and a lightweight, easy-to-use implementation can be found in the Python package \texttt{phayes}~\cite{phayes}.

\subsection{Parameter selection for $k$ and $\beta$}\label{subsec:PS}

We have yet to specify a method to select the circuit parameters $k$ and $\beta$.  Existing approaches \cite{Wiebe2016, OBrien2019, vandenBerg2021} typically use the heuristic $k = \big\lceil1.25/\sqrt{\text{Var}_{H}[\phi]}\big\rceil$ and sample $\beta$ randomly. Here we introduce an approach for deterministically selecting these parameters.

We first note that it is important to quantitatively assess our posterior uncertainty over the phase $\phi$, e.g. to determine a stopping criterion. This might be done by assessing the variance of the posterior distribution. However, for wrapped distributions, the standard variance $\text{Var}[\phi] =  \mathbb{E}_{p(\phi)}[(\phi - \mathbb{E}_{p(\phi)}[\phi])^2]$ is not well-behaved due to the misalignment of the squared error and the wrapped property \cite{Jupp2009, berry2002}. As such, two alternative circular metrics are used to assess concentration
\begin{align*}
    \text{Circular:}& &\text{Var}_{C}[\phi] &=  1 - |\mathbb{E}_{p(\phi)}[e^{i\phi}]| &&\in [0,1]. \\
    \text{Holevo:}& &\text{Var}_{H}[\phi] &=  |\mathbb{E}_{p(\phi)}[e^{i\phi}]|^{-2} -1 &&\in [0, \infty].
\end{align*}
The Holevo variance is often favoured due to it being unbounded above (akin to the standard variance). For sharply peaked distributions we have that $\text{Var}[\phi] \approx 2\text{Var}_C[\phi]$ and $\text{Var}[\phi] \approx \text{Var}_H[\phi]$.
\par
For both Fourier and von Mises distributions, the circular and Holevo variances are analytically tractable. Moreover, for the circular variance, we can calculate the expected post-measurement posterior circular variance to form a utility function
\begin{equation}
    U_C(k, \beta) 
    = 
    \sum_{m \in \{0,1\}} 
    -\text{Var}_C[\phi \mid m, k, \beta] p(m \mid k, \beta).
\end{equation}
A utility function \cite{Granade2012} describes some measure of information gained from an experiment as a function of the experiment parameters which we look to maximise, i.e. $k_r, \beta_r = \argmax_{k, \beta}U_C(k, \beta)$. In this specific case (circular variance utility and either Fourier or von Mises prior) we can analytically maximise the utility and therefore minimise the expected posterior variance (see Appendix~\ref{app:PS} for calculations and \texttt{phayes} \cite{phayes} for implementation).

\begin{figure}
    \centering
    \includegraphics[width=0.95\hsize]{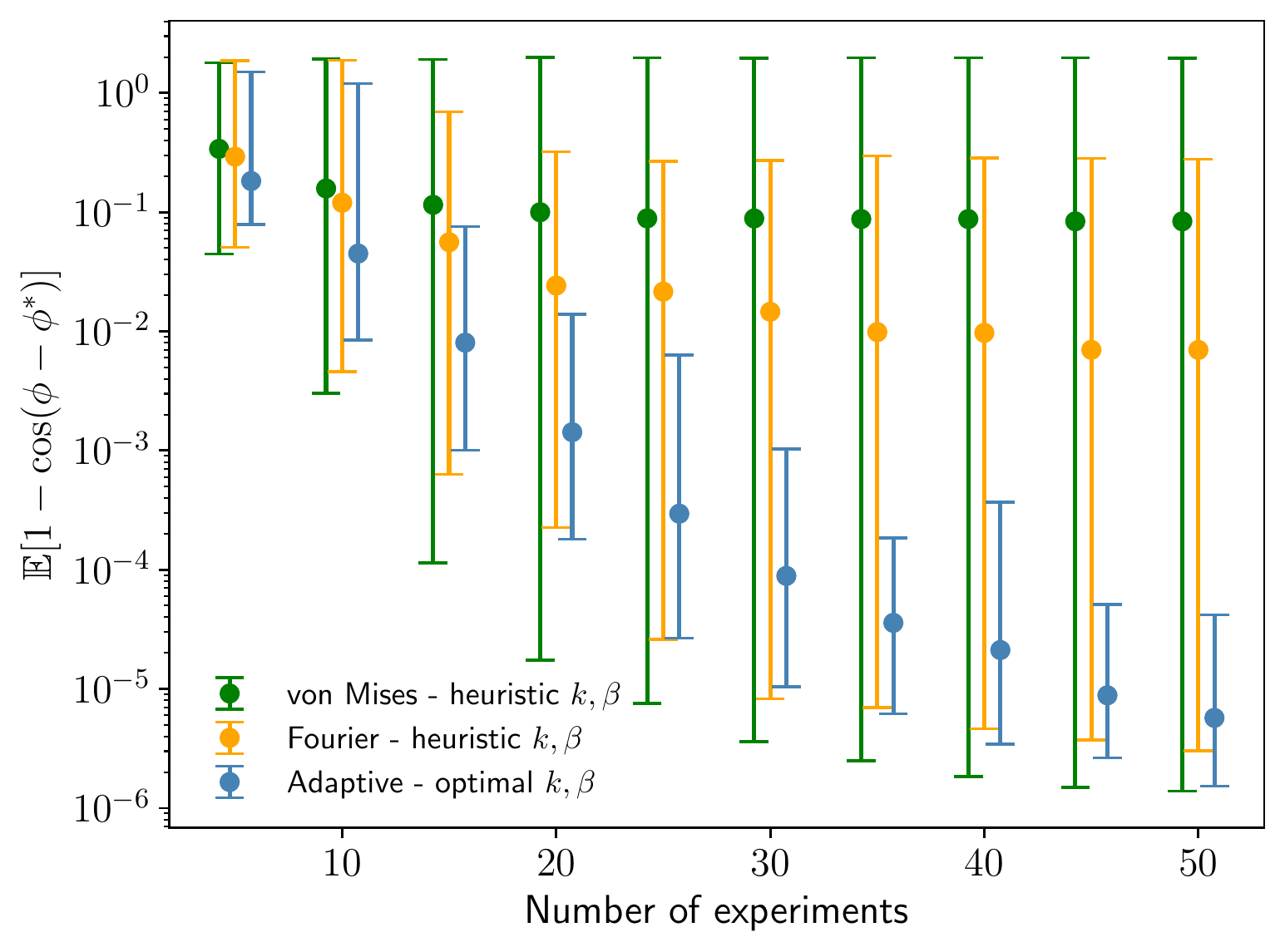}
    \caption{Comparison of Bayesian phase estimation inference and parameter selection methods. Simulations are noiseless and repeated over 100 different true phases $\phi^*$. Spots represent mean expected posterior cosine distance from the true phase and error bars display $\min$ and $\max$. The adaptive procedure converts to von Mises when the number of Fourier coefficients exceeds $J_\text{max} = 2000$. The selection of $k$ was capped at $k_\text{max} = 120$.
    }
    \label{fig:bpe_param_selection}
\end{figure}

We numerically compare Bayesian QPE approaches. 
For a given true phase $\phi^*\in[0,2\pi)$ to be estimated, the measurement outcomes are collected from synthetic experiments, where noiseless QPE circuits are simulated pseudo-randomly directly from the likelihood~\eqref{eq:likelihood}. This process is repeated over 100 different values $\phi^*$. 
Displayed in Fig.~\ref{fig:bpe_param_selection} is the mean with min/max error bars of the expected posterior cosine distance from the target angle $\phi^*$. 
In green, we make a von Mises approximation throughout and choose parameters according to the heuristic above, i.e., $k = \big\lceil1.25/\sqrt{\text{Var}_{H}[\phi]} \big\rceil$, $\beta$ sampled uniformly in $[0, 2\pi)$.
This approach is reminiscent of \cite{Wiebe2016} only with a von Mises distribution, rather than Gaussian, and Bayesian updates implemented via exact moment-matching rather than approximately with Monte Carlo. In orange, we update the true Fourier posterior again with the above heuristic parameter selection, matching the approach of \cite{OBrien2019}.
In blue, we apply the adaptive procedure described above and akin to \cite{vandenBerg2021} where we convert the exact posterior to von Mises after reaching $J_\text{max}=2000$ coefficients. In this adaptive implementation, we also apply the optimal parameter selection as described above and in Appendix~\ref{app:PS}. 
We observe that the optimal parameter selection ensures the Bayesian phase estimation is significantly more accurate, allowing the posterior to efficiently rule out incorrect modes and consistently hone in on the true phase.


\subsection{Noise-aware likelihood}

The likelihood in \eqref{eq:likelihood} assumes a noiseless implementation of the circuit \eqref{circ:hadamard}. In the presence of noise, it is more realistic to consider a noise-aware likelihood of the form~\cite{Wiebe2016}
\begin{equation}
\label{eq:noise_aware_likelihood}
    p(m \mid \phi, k, \beta) 
    = \frac{1 + (1-q)\cos(k\phi + \beta - m\pi)}{2},
\end{equation}
where $q$ is an error parameter due to decoherence.
%
The noise-aware likelihood \eqref{eq:noise_aware_likelihood} remains in a convenient Fourier form and still admits the same analytical Bayesian updates as \eqref{eq:likelihood} \cite{vandenBerg2021} as well as the aforementioned optimal parameter selection.

The error parameter $q$ is typically determined as a function of the quantum device specifications and the depth parameter $k$. Naturally, deeper circuits suffer from an increased error rate. It is therefore common in practice to consider some $k_\text{max}$ to limit the depth of the circuits and control the error rate.
For example, one finds $q=1-(1-p_2)^{N_\mathrm{2Q}}$ if the noisy QPE circuit $U_\mathrm{QPE}$ is modelled by the depolarizing channel that maps a density matrix $\rho$ to 
\begin{equation}
\label{eq:depolarizing}
    (1-p_2)^{N_\mathrm{2Q}}U_\mathrm{QPE}\rho U^\dag_\mathrm{QPE} + (1-(1-p_2)^{N_\mathrm{2Q}})\frac{I}{2^n},
\end{equation}
with the depolarizing error rate $p_2$ of two-qubit gates and the number of two-qubit gates $N_\mathrm{2Q} = \calO(k)$ in the circuit.

\section{Error detection with $\llbracket6,4,2\rrbracket$ code}
\label{sec:QED}

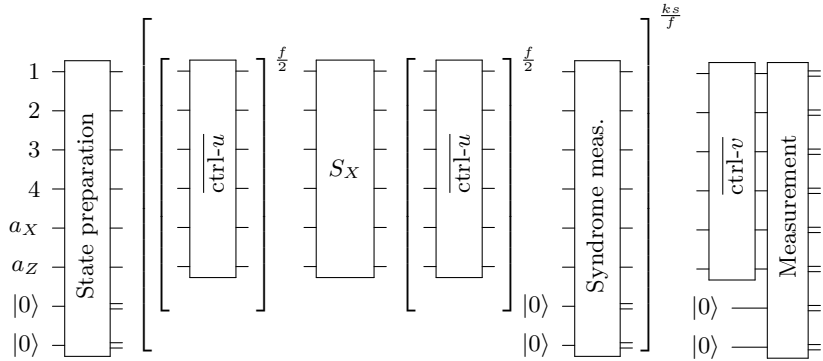
\begin{figure*}
\begin{align}
\nonumber
    \begin{array}{c}
        \Qcircuit @C=.5em @R=0.7em {
            \\
            \\
            \lstick{1}
            & \multigate{7}{\rotatebox[origin=l]{90}{State preparation}}
            & \qw
            \\
            \lstick{2}
            & \ghost{\rotatebox[origin=l]{90}{State preparation}}
            & \qw
            \\
            \lstick{3}
            & \ghost{\rotatebox[origin=l]{90}{State preparation}}
            & \qw
            \\
            \lstick{4}
            & \ghost{\rotatebox[origin=l]{90}{State preparation}}
            & \qw
            \\
            \lstick{a_X}
            & \ghost{\rotatebox[origin=l]{90}{State preparation}}
            & \qw
            \\
            \lstick{a_Z}
            & \ghost{\rotatebox[origin=l]{90}{State preparation}}
            & \qw
            \\
            \lstick{\ket{0}}
            & \ghost{\rotatebox[origin=l]{90}{State preparation}}
            & \cw
            \\
            \lstick{\ket{0}}
            & \ghost{\rotatebox[origin=l]{90}{State preparation}}
            & \cw
        }
    \end{array}
    \left[
    \left[
    \begin{array}{c}
        \Qcircuit @C=.5em @R=.7em {
            & \multigate{5}{\rotatebox[origin=l]{90}{$\overline{\text{ctrl-}u}$}}
            & \qw
            \\
            & \ghost{\rotatebox[origin=l]{90}{$\overline{\text{ctrl-}u}$}}
            & \qw
            \\
            & \ghost{\rotatebox[origin=l]{90}{$\overline{\text{ctrl-}u}$}}
            & \qw
            \\
            & \ghost{\rotatebox[origin=l]{90}{$\overline{\text{ctrl-}u}$}}
            & \qw
            \\
            & \ghost{\rotatebox[origin=l]{90}{$\overline{\text{ctrl-}u}$}}
            & \qw
            \\
            & \ghost{\rotatebox[origin=l]{90}{$\overline{\text{ctrl-}u}$}}
            & \qw
            \\
            \\
            \\
        }
    \end{array}
    \right]^{\frac{f}{2}}
    \begin{array}{c}
        \Qcircuit @C=.5em @R=0.7em {
            & \multigate{5}{S_X}
            & \qw
            \\
            & \ghost{S_X}
            & \qw
            \\
            & \ghost{S_X}
            & \qw
            \\
            & \ghost{S_X}
            & \qw
            \\
            & \ghost{S_X}
            & \qw
            \\
            & \ghost{S_X}
            & \qw
            \\
            \\
            \\
        }
    \end{array}
    \left[
    \begin{array}{c}
        \Qcircuit @C=.5em @R=.7em {
            & \multigate{5}{\rotatebox[origin=l]{90}{$\overline{\text{ctrl-}u}$}}
            & \qw
            \\
            & \ghost{\rotatebox[origin=l]{90}{$\overline{\text{ctrl-}u}$}}
            & \qw
            \\
            & \ghost{\rotatebox[origin=l]{90}{$\overline{\text{ctrl-}u}$}}
            & \qw
            \\
            & \ghost{\rotatebox[origin=l]{90}{$\overline{\text{ctrl-}u}$}}
            & \qw
            \\
            & \ghost{\rotatebox[origin=l]{90}{$\overline{\text{ctrl-}u}$}}
            & \qw
            \\
            & \ghost{\rotatebox[origin=l]{90}{$\overline{\text{ctrl-}u}$}}
            & \qw
            \\
            \\
            \\
        }
    \end{array}
    \right]^{\frac{f}{2}}
    \begin{array}{c}
        \Qcircuit @C=.5em @R=.7em {
            \\
            \\
            &
            & \multigate{7}{\rotatebox[origin=l]{90}{Syndrome meas.}}
            & \qw
            \\
            &
            & \ghost{\rotatebox[origin=l]{90}{Syndrome meas.}}
            & \qw
            \\
            &
            & \ghost{\rotatebox[origin=l]{90}{Syndrome meas.}}
            & \qw
            \\
            &
            & \ghost{\rotatebox[origin=l]{90}{Syndrome meas.}}
            & \qw
            \\
            &
            & \ghost{\rotatebox[origin=l]{90}{Syndrome meas.}}
            & \qw
            \\
            &
            & \ghost{\rotatebox[origin=l]{90}{Syndrome meas.}}
            & \qw
            \\
            & \lstick{\ket{0}}
            & \ghost{\rotatebox[origin=l]{90}{Syndrome meas.}}
            & \cw
            \\
            & \lstick{\ket{0}}
            & \ghost{\rotatebox[origin=l]{90}{Syndrome meas.}}
            & \cw
        }
    \end{array}
    \right]^{\frac{ks}{f}}
    \begin{array}{c}
    \vspace{-2em}
        \Qcircuit @C=.5em @R=0.7em {
            & \multigate{5}{\rotatebox[origin=l]{90}{$\overline{\text{ctrl-}v}$}}
            & \multigate{7}{\rotatebox[origin=l]{90}{Measurement}}
            & \cw
            \\
            & \ghost{\rotatebox[origin=l]{90}{$\overline{\text{ctrl-}v}$}}
            & \ghost{\rotatebox[origin=l]{90}{Measurement}}
            & \cw
            \\
            & \ghost{\rotatebox[origin=l]{90}{$\overline{\text{ctrl-}v}$}}
            & \ghost{\rotatebox[origin=l]{90}{Measurement}}
            & \cw
            \\
            & \ghost{\rotatebox[origin=l]{90}{$\overline{\text{ctrl-}v}$}}
            & \ghost{\rotatebox[origin=l]{90}{Measurement}}
            & \cw
            \\
            & \ghost{\rotatebox[origin=l]{90}{$\overline{\text{ctrl-}v}$}}
            & \ghost{\rotatebox[origin=l]{90}{Measurement}}
            & \cw
            \\
            & \ghost{\rotatebox[origin=l]{90}{$\overline{\text{ctrl-}v}$}}
            & \ghost{\rotatebox[origin=l]{90}{Measurement}}
            & \cw
            \\
            &\lstick{\ket{0}}
            & \ghost{\rotatebox[origin=l]{90}{Measurement}}
            & \cw
            \\
            &\lstick{\ket{0}}
            & \ghost{\rotatebox[origin=l]{90}{Measurement}}
            & \cw
        }
    \end{array}
\end{align}
\caption{\label{fig:logical_circ} A sketch of the entire encoded QPE circuit. It starts with the state preparation with two ancillary qubits at the bottom.
Then, the logical controlled unitary operations, $\overline{\text{ctrl-}u}$, are sequentially applied with the syndrome measurements interweaved.
In every $f$ step of time evolution, a syndrome measurement is performed and executions are discarded upon detecting errors.
A stabilizer $S_X$ is inserted in each block to alleviate physical memory errors.
The $\overline{\text{ctrl-}v}$ operation is applied right before the measurement (the logical $R_Z(\beta)$ is absorbed into $\overline{\text{ctrl-}v}$).
After all the logical unitary operations are performed, the final fault-tolerant measurement is made to read off logical Pauli expectation values.}
\end{figure*}

Aiming at hardware demonstration of the aforementioned QPE protocol, we consider how to protect our quantum circuit from errors caused by noisy quantum hardware. 
Quantum error detection (QED) codes provide such protection with limited quantum resources by discarding erroneous executions of quantum circuits. For instance, the $\llbracket n+2,n,2\rrbracket$ code for even $n$ is a stabilizer code whose code space is stabilized by $X^{\otimes n+2}$ and $Z^{\otimes n+2}$~\cite{Steane:1996va,Gottesman:1997qd,knill2004scheme,knill2004threshold} (dubbed Iceberg code in~\cite{Self2022}).
In this work, we employ the code with $n=4$, i.e. $\llbracket6,4,2\rrbracket$ code, to encode four logical qubits.
The logical qubits and two redundant physical qubits, denoted by  $L:=\{1,2,3,4\}$ and $A:=\{a_X,a_Z\}$, form the six-qubit code on $T:=L\cup A$. 
More concretely, the three-qubit QPE circuit, with one dummy qubit appended, is encoded into six physical qubits.
We introduce two additional ancillary qubits to carry out fault-tolerant state preparation, syndrome extractions, and final measurement, leading to eight qubits used in total.
We denote the two stabilizers by $S_X:=\bigotimes_{i\in T}X_i$ and $S_Z:=\bigotimes_{i\in T}Z_i$, and their simultaneous eigenstates associated with the eigenvalue +1 define the four-qubit logical space. 
Reading out $-1$ upon measuring $\{S_X,S_Z\}$ signals the errors that do not commute with the stabilizer operators, and thus such circuit executions are discarded. The undetectable errors by stabilizer (syndrome) measurements lead to logical errors disturbing the encoded system.

The encoded quantum states are manipulated by logical Pauli operators
\begin{align}
\label{eq:logical}
     \bar{X}_i:=X_iX_{a_X}, 
     \quad 
     \bar{Z}_i:=Z_iZ_{a_Z}, 
     \quad 
     \text{for } i\in L,
\end{align}
which commute with the stabilizers and obey $\{\bar{X}_i,\bar{Z_i}\}=[\bar{X}_i,\bar{Z}_j]=0$ for $i\neq j$.
All the logical operations are compiled to the universal logical gate set, which, in the form of physical gates, is given by\footnote{
For example, the physical operation $R_{X_iX_j}(\theta)$ corresponds to the following logical operations,
\begin{align}
\left\{
\begin{array}{ll}
    R_{\bar{X}_i}(\theta)
    & (i\in L,\ j= a_X),
    \\
    R_{\bar{X}_i\bar{X}_j}(\theta)
    & (i ,j \in L),
    \\
    R_{\bar{X}_k\bar{X}_l\bar{X}_m}(\theta)
    & (\{i,k,l,m\} = L,\ j=a_Z),
    \\
    R_{\bar{X}_k\bar{X}_l\bar{X}_m\bar{X}_n}(\theta)
    & (\{k,l,m,n\} = L,\ i=a_X,\ j=a_Z).
\end{array}
\right.
\end{align}
}~\cite{Self2022}
\begin{equation}
\label{eq:universal_gates}
    \{R_{P_iP_j}(\theta)\mid i, j\in T,\ i\neq j,\ P\in\{X, Y, Z\}\},
\end{equation}
where we have defined a Pauli exponential operator $R_P(\theta):=e^{-i\theta P/2}$ for a rotation angle $\theta$ and a Pauli operator~$P$.

Figure~\ref{fig:logical_circ} shows the entire structure of the encoded QPE circuit.
See Appendix~\ref{app:QED} for explicit compilation of each component.
For concreteness, we let the operator, $U^k$, in the QPE circuit an approximation to $e^{-i Hkt}$ with Hamiltonian~\eqref{eq:hamiltonian_h2} and $t\in\mathbb{R}$ by the Lie-Trotter first-order product formula~\cite{Lloyd1996},
\begin{equation}
\label{eq:product}
\begin{aligned}
    U^{k}(t)
    &:=
    \left(
        e^{-i\frac{h_{1}t}{s}Z_{1}}
        ~
        e^{-i\frac{h_{2}t}{s}Z_{2}}
        ~
        e^{-i\frac{h_{3}t}{s}Y_{1}Y_{2}}
    \right)^{ks}
    \\
    &\quad\times
    e^{-\im h_{4}ktZ_{1}Z_{2}}
    e^{-\im h_{5}ktI}
    \\
    &= e^{-i Hkt}+\calO(kt^2/s).
\end{aligned}
\end{equation}
Here, the integer $s$ is the number of discretized time steps.
Then, the controlled version of the unitary $u:=e^{-i\frac{h_{1}t}{s}Z_{1}} e^{-i\frac{h_{2}t}{s}Z_{2}} e^{-i\frac{h_{3}t}{s}Y_{1}Y_{2}}$ is compiled to logical operators, which is denoted by $\overline{\text{ctrl-}u}$ in Fig.~\ref{fig:logical_circ}. The remaining part $v:=e^{-\im h_{4}ktZ_{1}Z_{2}} e^{-\im h_{5}ktI}$ is similarly compiled to logical operators $\overline{\text{ctrlu-}v}$ and applied at the end because it commutes with $u$ and does not require the Lie-Trotter decomposition.

The encoded circuit starts with encoding the initial state. The two extra ancillary qubits are dedicated to performing the fault-tolerant initialization.
We split the logical operation $(\overline{\text{ctrl-}u})^{ks}$ into $\lfloor ks/f\rfloor$ blocks. 
Application of each block is followed by a syndrome measurement to inspect whether the encoded state is stabilized by $\{S_X, S_Z\}$. As soon as one syndrome measurement is read off as $-1$, the circuit execution is aborted.
Furthermore, we insert a stabilizer $S_X$ in the middle of each block, which acts on the code space as the identity.
However, it suppresses the coherent physical errors in the form of a single-qubit $Z$ rotation accumulating in time (Appendix~\ref{app:memory}).
After all the logical unitary operations are performed, we make a final measurement to ensure that both stabilizers are measured to 1 and read off logical Pauli expectation values.

We remark on some features of the code exploiting the capability of Quantinuum's H1-1 trapped-ion computer~\cite{Ryan-Anderson2022,Self2022}. 
The state preparation, syndrome measurements, and projective measurement are performed in a fault-tolerant manner~\cite{gottesman2016quantum,Chao2017FaulttolerantQC,Chao2018,Linke2017}.
While non-fault-tolerantly implemented, a logical operator $R_{P_iP_j}(\theta)$~\eqref{eq:universal_gates} is, up to single-qubit Clifford gates, compiled to a single M{\o}lmer-S{\o}rensen (MS) gate, $R_{Z_iZ_j}(\theta)$, which is natively implemented on the trapped-ion computer with the gate infidelity $\sim 2\times 10^{-3}$~\cite{Molmer1999,H1datasheet}.
As such, Quantinuum's high-fidelity MS gate operations combined with the all-to-all connectivity is expected to lead to logical circuit executions with a low logical error rate.\footnote{Fault-tolerant implementations of logical gates are proposed~\cite{Gottesman:1997qd,Chao2017FaulttolerantQC}. We do not pursue such protocols here to avoid extra overheads.} 
Lastly, the Quantinuum device is equipped with {\it conditional exit}, a functionality to immediately abort the circuit execution conditioned on classical registers. Thanks to this feature, one can save runtime by discarding calculations as soon as a syndrome measurement detects errors without running the remaining calculations in vain.


\section{Experiments}
\label{sec:experiment}

We experimentally demonstrate the Bayesian QPE protocol adopting the parameter selection introduced in Sec.~\ref{subsec:PS} with and without the $\llbracket 6,4,2\rrbracket$ error detection code on the Quantinuum H1-1 quantum computer,
after calibrating the individual QPE circuit~\eqref{circ:hadamard}. We calculate the ground state energy of the Hamiltonian~\eqref{eq:hamiltonian_h2} with the coefficients $(h_{1},h_{2},h_{3},h_{4},h_{5})=(-0.3980,-0.3980,-0.1809,0.0112,-0.3322)$ hartree.
It describes the equilibrium geometry of the hydrogen molecule with the inter-atomic distance $R_{\mathrm{HH}} = 0.73486$~{\AA}.
The experimental results shown in this section are calculated using the product formula~\eqref{eq:product} with $t=0.1\pi$ and $s=1$ in atomic unit of time. 
This yields $-2.0 \times 10^{-4}$ hartree of systematic error.
The estimated energy $E$ is related to the estimated phase $\phi$ by $E=-\phi/t$.

The present experiments rely on the following software packages. 
The quantum circuits are prepared with \texttt{pytket} v1.13.2~\cite{Sivarajah2020-jg},
and executed with \texttt{pytket-quantinuum} v0.15.0.
We use \texttt{InQuanto} v2.1.1~\cite{inquanto} and its interface to \texttt{pyscf}
v2.2.0~\cite{Sun2018-gy} to calculate the coefficients of the spin Hamiltonian \eqref{eq:hamiltonian_h2} on classical computers. The classical pre and post-processing for parameter selection and Bayesian inference are handled by \texttt{phayes} v0.0.3~\cite{phayes}.

\subsection{Calibration of QPE}
\label{subsec:benchmark}

We start by calibrating the performance of individual QPE circuits.
The benchmark informs us of the quantitative relationship between the circuit depth $k$ and noise level $q$, which limits the precision of phase estimation.
It also showcases that the QED primitive can suppress hardware noise to enable experiments with deeper circuits.

To this end, we evaluate the probability of obtaining the measurement outcome ``0'' from the QPE circuit~\eqref{circ:hadamard} in comparison with what one would expect from a noiseless experiment.
In order to reduce the source of systematic errors, we use the exact eigenstate $\ket{\phi_0}$ of the operator $U$ with the eigenvalue $\phi_0$, giving an approximation to the ground state energy of $H$ up to the error of product formula~\eqref{eq:product}.


The unencoded circuit~\eqref{circ:hadamard} (see \ref{circ:QPE}-\ref{eq:ctrlV} for details) contains $N_\text{2Q}^{\text{(u)}}$ two-qubit (2Q) gates with\footnote{A ctrl-$u$ uses 5 2Q gates, a ctrl-$v$ has 4 2Q gates, and the preparation of Hartree--Fock (exact ground) state uses 0 (1) 2Q gates.}
\begin{equation}
\label{eq:2Q_count_unenc}
    N_\text{2Q}^{\text{(u)}}(k) = 5k + 4 + \Delta_\text{init},
\end{equation}
where $\Delta_\text{init}=0\ (1)$ when the input state is the Hartree-Fock state (exact eigenstate).
For the experiment with QED, we use the encoded circuit (Fig.~\ref{fig:logical_circ}), which contains $N_\text{2Q}^{\text{(e)}}$ 2Q gates,\footnote{A $\overline{\text{ctrl-}u}$ contains 6 2Q gates, each syndrome measurement uses 12 2Q gates, the preparation of encoded Hartree-Fock (exact ground) state uses 9 (14) 2Q gates, the $\overline{\text{ctrl-}v}$ uses 3 2Q gates, and 8 2Q gates are used in the final measurement.}
\begin{equation}
\label{eq:2Q_count_enc}
    N_{\text{2Q}}^{\text{(e)}}(k)
    =
    6k
    + 12\left\lfloor \frac{k}{f} \right\rfloor
    + 20
    + 5 \Delta_\text{init}.
\end{equation}
The frequency of syndrome measurements is fixed to $f=8$ throughout the encoded experiments presented in this section.

\begin{figure}
    \centering
    \includegraphics[width=0.90\hsize]{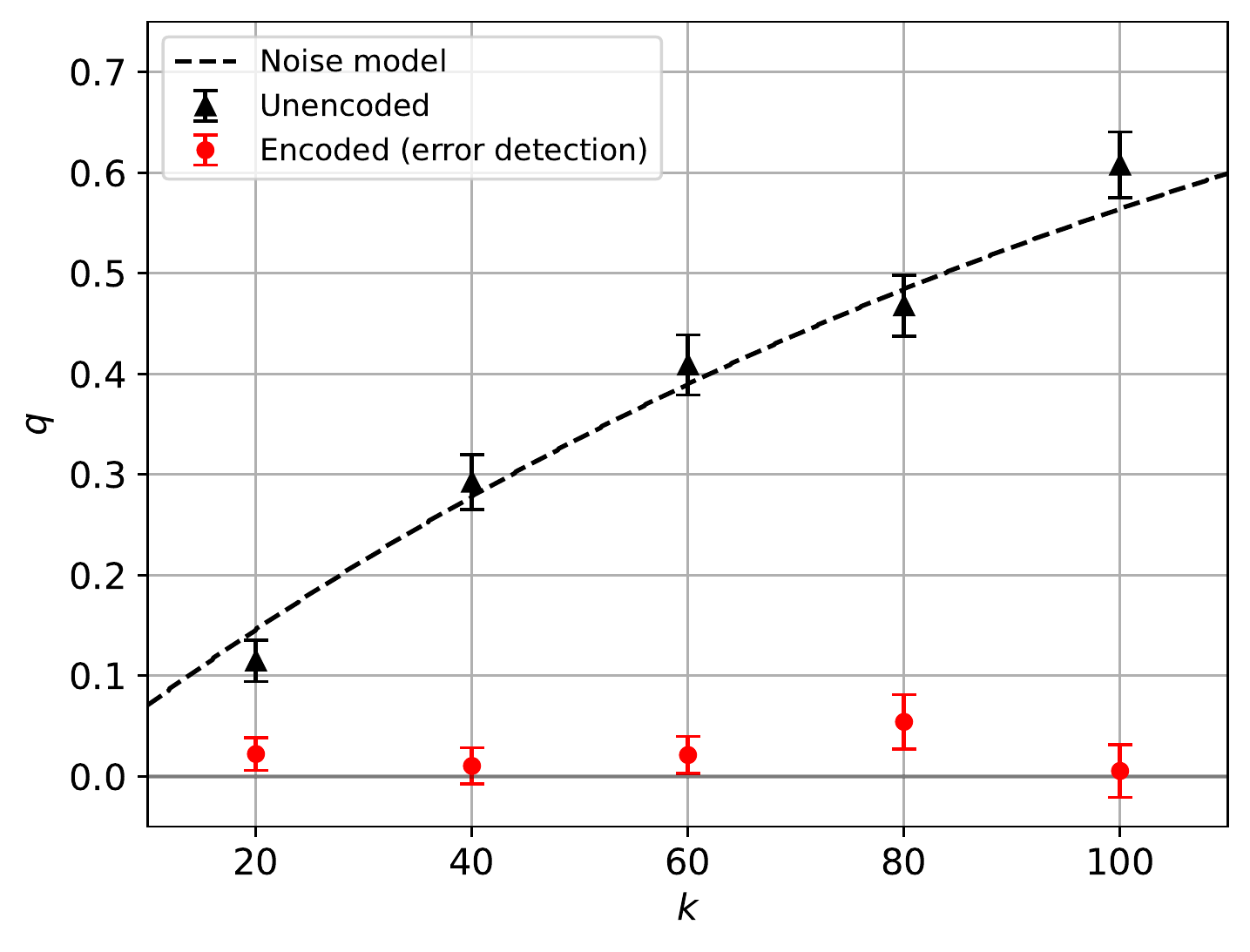}
    \caption{
         Parameter $q$ obtained by fitting the experimental results with the function~\eqref{eq:fit_func} for the unencoded (black triangles) and encoded (red circles) cases. The error bars represent the statistical uncertainties due to a finite number of measurements. The black dashed curve shows the function~\eqref{eq:model_p2} with $p_{2} = 1.6 \times 10^{-3}$.
    }
    \label{fig:error_rate}
\end{figure}

For calibration the integer $k$ is chosen from $\{20, 40, 60, 80, 100\}$, and for each $k$,
we pick $\beta$ from the four sample points
$\{
    -k\phi_{0}-\pi,
    -k\phi_{0}-\pi/2,
    -k\phi_{0},
    -k\phi_{0}+\pi/2
\}$.
We run 500 shots of calculations for each parameter set $\{k,\beta\}$ and fit the experimental data with the function,
\begin{equation}
\label{eq:fit_func}
    f(k, \beta; q, \omega) 
    = \frac{1 + (1-q)\cos(k(\phi_0-\omega) + \beta)}{2},
\end{equation}
to extract the parameters $q$ and $\omega$ that characterize the amplitude damping
and the shift in the estimated phase, respectively.
In the absence of hardware noise, both $q$ and $\omega$ vanish to recover the noiseless likelihood~\eqref{eq:likelihood}.
The experimental results with and without QED are shown in Fig.~\ref{fig:error_rate}.
The $k$-dependence of $q$ obtained from the experiments with unencoded circuits is captured by the following function,
\begin{equation}
    \label{eq:model_p2}
    q(k; p_{2}) = 1 - (1 - p_{2})^{N_\mathrm{2Q}^{\text{(u)}}(k)},
\end{equation}
with $p_2=1.6\times10^{-3}$.
Equation~\eqref{eq:model_p2} is derived assuming the depolarizing noise model~\eqref{eq:depolarizing}.
The parameter $p_2$ gives a rough estimate of average two-qubit gate infidelity, whose order of magnitude matches the value reported in~\cite{H1datasheet}.
While $q$ rapidly grows as the circuit depth $k$ increases in the unencoded case, it is well suppressed in the encoded case thanks to the protection by the $\llbracket6,4,2\rrbracket$ code.
As presented in Appendix~\ref{app:gate_error_mitigation}, for both unencoded and encoded benchmarks, the phase shifts $\omega$ are of order $10^{-3}$ (Fig.~\ref{fig:benchmark_omega}).
In this work, we omit $\omega$ in the noise-aware likelihood function~\eqref{eq:noise_aware_likelihood} in the Bayesian QPE and leave it as future work to explore coherent error mitigation techniques~\cite{Lider2014,Suter2016,Wallman2016, Cai:2022rnq, Gu:2022ziq}.

\begin{figure}
    \centering
    \includegraphics[width=0.90\hsize]{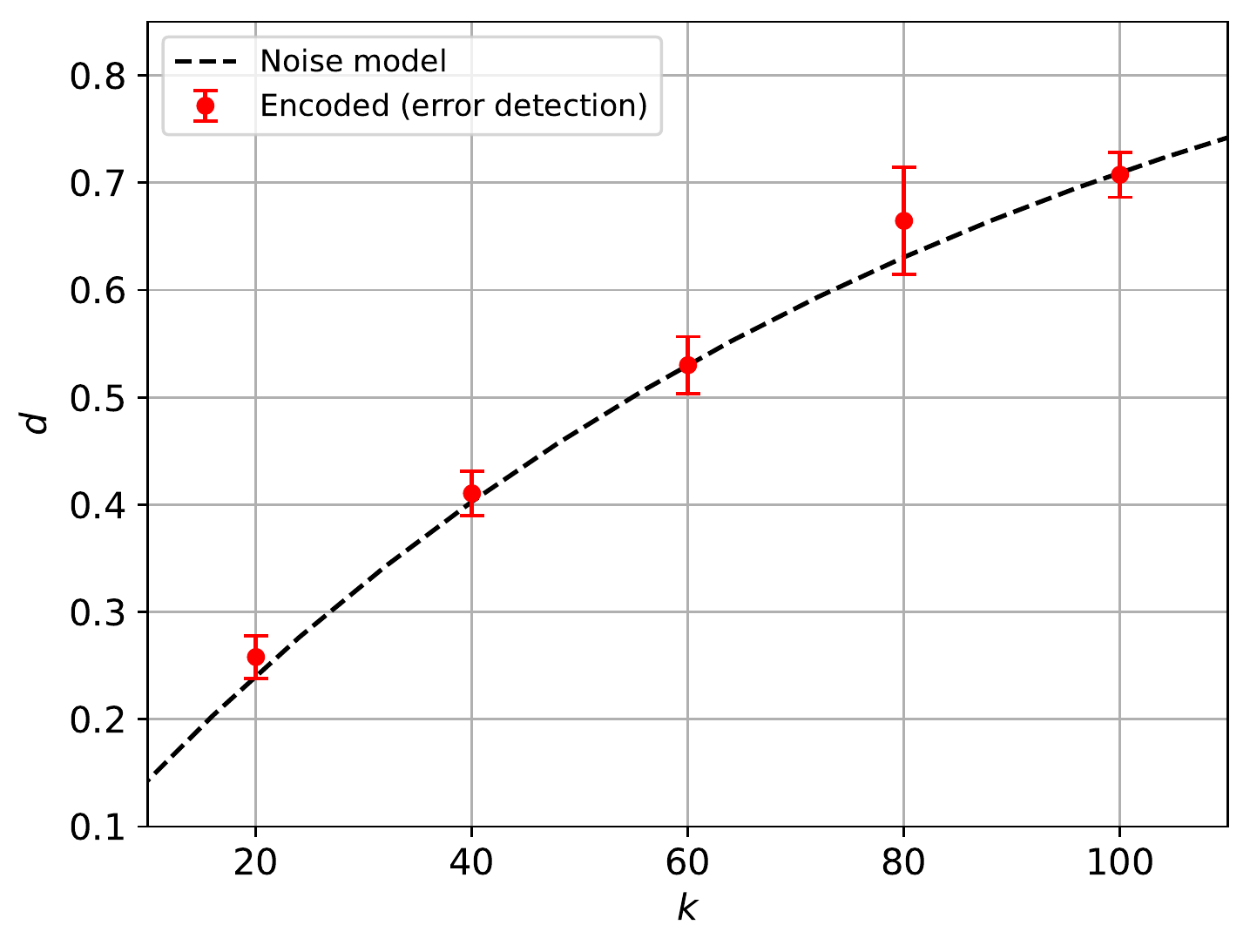}
    \caption{
        The discard rates of the encoded circuit calibrations with respect to $k$ are represented by red data points. The dashed line shows~\eqref{eq:model_p2} with $p_{2}=1.6\times 10^{-3}$.
    }
    \label{fig:survival_rate}
\end{figure}

The suppression of error rate with QED is accomplished at the cost of discarding faulty computations that are detected by syndrome measurements.
The discard rate $d$, the ratio of the number of discarded circuits to the total number of executed circuits, with respect to the circuit depth $k$ is shown in Fig.~\ref{fig:survival_rate}.
The same error model~\eqref{eq:model_p2},
\begin{equation}
    \label{eq:model_p2_d}
    d(k; p_{2}) = 1 - (1 - p_{2})^{N_\mathrm{2Q}^{\text{(e)}}(k)},
\end{equation}
with $p_2=1.6\times10^{-3}$ recovers the experimentally obtained discard rate.
It is, however, not obvious that the discard rate $d$ is explained by~\eqref{eq:model_p2_d}, which is based on the depolarizing noise model. 
The increase of $q$ of \eqref{eq:fit_func} in the unencoded case mainly stems from incoherent errors. On the other hand, the errors detected by the QED contain all the single-qubit errors, and thus, the discard rate is generally influenced by a wider class of errors.
However, we inserted an $X$-stabilizer $S_X$ in the middle of each block in the encoded circuit~(Fig.~\ref{fig:logical_circ}) to suppress the dominant source of coherent errors. 
While $S_X$ acts as the identity on logical states, we confirm that the stabilizer insertion suppresses coherent error, which in turn reduces $d$ and $q$ as further discussed in Appendix~\ref{app:memory} (see Fig.~\ref{fig:error_discard_Sx}).
Consequently, the discard rate in the current encoded experiments is dominated by incoherent errors and is well captured by the depolarizing model.

\subsection{Bayesian QPE}
\label{subsec:bayesian_qpe}

\begin{figure*}
    \centering
    \includegraphics[width=0.95\hsize]{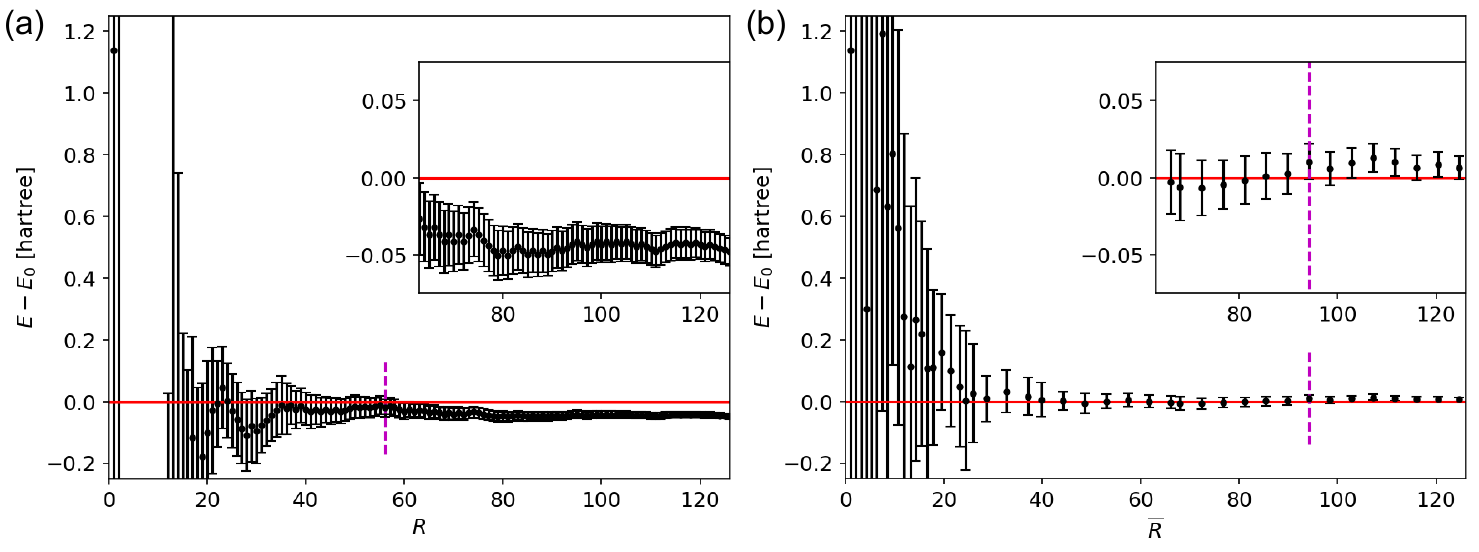}
    \caption{
        Estimated energies plotted against the number of experiments $R$ for the unencoded QPE (a) and against the rescaled number of experiments $\overline{R}$ for the encoded experiments (b).
        The black circles represent the estimated energies with the error bars representing $\sqrt{\text{Var}_H[E]}$. 
        The purple dashed lines indicate when the distributions are converted from Fourier to von Mises representations.
    }
    \label{fig:bqpe_energy}
\end{figure*}

Having calibrated the performance of QPE circuits, we turn to the Bayesian QPE experiments with and without QED.
As an input state $\ket{\phi}$ of QPE, we prepare the Hartree-Fock state $\ket{00}$.
The Hartree-Fock state provides the mean-field approximation to the ground state of $H$ and is commonly used in computational chemistry to evaluate the electronic correlation energy.
For the Hamiltonian given at the beginning of the section, the fidelity between the Hartree-Fock state and the exact ground state is $0.981$.
Therefore, it is not expected that our demonstration of Bayesian QPE is significantly influenced by contamination from the other eigenstates. In Appendix~\ref{app:noiseless_sim}, we numerically confirm this by conducting noiseless QPE simulations.

The number of Fourier coefficients in~\eqref{eq:fourier_posterior} is set to $J_{\text{max}} = 2000$, and the maximum depth is specified by $k_{\text{max}} = 120$.
The number of 2Q gates in the unencoded QPE circuit is up to $N_{\mathrm{2Q}}^{\text{(u)}}(k_\text{max})=604$, while it is as many as $N_{\mathrm{2Q}}^{\text{(e)}}(k_\text{max})=920$ in the encoded circuit with the expected discard rate $d(k_\text{max};p_2)=0.77$. We use $p_2=1.6\times10^{-3}$ throughout the subsection.


The estimated energy $E$ is plotted against the number of circuit executions in Fig.~\ref{fig:bqpe_energy}.
To take into account the discarded experiments in the QPE with QED, we use the rescaled number of experiments given by
\begin{equation}
    \overline{R}
    := \sum_{r=1}^{R} n_\text{shots}(k_r),
    \quad
    n_\text{shots}(k_r) := \frac{1}{1 - d(k_r;p_2)},
\end{equation}
where $k_r$ stands for the depth of the circuit in the $r$-th round of the Bayesian QPE experiment, $n_\text{shots}(k_r)$ is an average number of experiments until the circuit of depth $k_r$ is executed without any error detected, and $R$ is the number of Bayesian updates~\eqref{eq:posterior}. 
%
The calculated energies $E$ with the noise-aware likelihood is $-1.185\pm0.009$~hartree after 125 Bayesian updates of distributions for the exact ground state energy $E_0=-1.1375$~hartree.
On the other hand, with the QED, we find $E=-1.131\pm 0.007$~hartree after 44 Bayesian updates.
There are two main sources of the systematic error observed in the unencoded experiment~[Fig.~\ref{fig:bqpe_energy}(a)]. 
One is the coherent errors that our noise-aware likelihood function~\eqref{eq:noise_aware_likelihood} fails to capture.
It implies that, with the use of the noise-aware likelihood based on better characterizations of coherent errors, the precision can be further improved.
This also shows the benefit of using the QED, which does not require aggressive characterization of hardware noise.
The other source is that the unencoded experiment has less informative measurements (larger error parameter $q$ of \eqref{eq:noise_aware_likelihood}), and therefore, the adaptive Bayesian inference scheme necessarily converts to a von Mises approximation earlier than in the encoded case for the same $J_\text{max}$. This leads to a higher probability of algorithmic error as observed in the synthetic experiments~(Fig.~\ref{fig:bpe_param_selection}).

Finally, we remark that in the encoded experiments the conditional exit operation provided by the H1 computer is used to reduce the total runtime. The feature allows one to abort the calculation as soon as the error is detected by a syndrome measurement.
For example, provided the encoded QPE circuits of depth $k = \{40, 80, 120\}$, the ratios of the average number of 2Q gates before the exit to the total number of 2Q gates are given by $\{0.83, 0.67, 0.56\}$.
This ratio becomes 0.60 if we accumulate the average and total number of 2Q gates along the Bayesian QPE experiment.
It should be, however, noted that whether the reduction of depth by the conditional exit leads to the reduction of actual runtime depends on various other overheads.


\section{Conclusion}
\label{sec:conclusion}

We conducted QPE experiments on the hydrogen molecule Hamiltonian to estimate its ground state energy using the Quantinuum H1-1 trapped-ion quantum device.
To the best of our knowledge, it is the first experiment where the QPE on logical qubits is applied to a quantum chemistry problem. We note that~\cite{Urbanek2020,Zhang2022} addressed the same problem using the variational quantum eigensolver and the $\llbracket4,2,2\rrbracket$ code.
We obtained good agreement with the exact ground state energy, which is due to both the algorithmic advances we have applied and the capability of the H1-1 device.

To make the best use of the limited circuit volume that can be reliably run, we employed the Bayesian approach to the QPE with an enhanced parameter selection which we have demonstrated to have significantly improved robustness properties.
Furthermore, we tested two approaches to cope with the hardware noise and imperfections.
In the first approach, we adopted the noise-aware likelihood function for Bayesian update, assuming that the depolarizing channel models the dominant source of errors.
We calculated the energy with the error $0.048\pm0.009$~hartree using the QPE circuit containing up to 604 two-qubit gates.
In the second approach, the QPE circuit is encoded with the $\llbracket 6,4,2\rrbracket$ quantum error detection code. In the encoded circuit, a simple coherent error mitigation via $X$-stabilizer insertions is applied to achieve lower discard and logical error rates.
With the encoded circuits, the ground state energy was estimated within $0.006\pm0.007$~hartree of the exact value. The largest circuit contains 920 physical two-qubit gates with the discard rate $\sim 77\%$.
While our experimental results show a slightly better performance with the QED, there is room for improvement in both approaches.
For instance, more elaborate characterizations of hardware noise and error mitigation strategies can be deployed~\cite{Temme2017,Li2017,Cai:2022rnq}.
This allows one to alleviate the coherent errors which were omitted in our noise-aware likelihood function for the unencoded experiment~\cite{Lider2014,Suter2016,Wallman2016,Gu:2022ziq}. The error mitigation also helps the encoded experiment by reducing the discard rate and logical error rate.

In the present work, we focused on a specific QPE algorithm, taking advantage of the capability of the device, for which the error detection code is highly optimized. In the future, it will be interesting to investigate other types of QPE algorithms such as the one suggested in~\cite{Lin2022,Wan2022} to further improve the performance of ground state energy estimation towards its application to more practical problems.

\section*{Data availability} 

The data presented in this paper are available from the corresponding author upon reasonable request.

\section*{Code availability}

The codes to reproduce the data presented in this paper are available from the corresponding author upon reasonable request.

\section*{Acknowledgments} 

We thank
David Amaro,
Hussain Anwar,
Charles H. Baldwin,
Marcello Benedetti,
Yi-Hsiang Chen,
Silas Dilkes,
David Hayes,
Steven Herbert,
Dominic G. Lucchetti,
Karl Mayer,
Brian Neyenhuis,
Matthias Rosenkranz,
Ciar\'{a}n Ryan-Anderson,
Christopher N. Self,
Andrew Tranter,
Ifan Williams,
and
Yasuyoshi Yonezawa
for useful discussions and comments on an earlier version of the manuscript.

\section*{Author contributions} 

K.Y., S.D. and Y.K. conceived and designed the study. 
K.Y. and S.D. performed numerical studies. 
K.Y. carried out experiments on quantum hardware.
All authors analysed the data, interpreted the results, and wrote the manuscript.

\section*{Competing interest} 

The authors declare no competing interests.


\bibliography{bib_prxq}

\newpage
\clearpage
\appendix
\onecolumngrid

\begin{center}
\Large Supplementary information
\end{center}

\renewcommand\thefigure{\thesection \arabic{figure}}
\renewcommand\thetable{\thesection \arabic{table}}

\section{Details on Bayesian phase estimation}
\label{app:BPE}
\setcounter{figure}{0} 

\subsection{Bayesian updates}\label{app:BPE_updates}

Recall the general phase estimation likelihood \eqref{eq:noise_aware_likelihood}
\begin{equation*}
    p(m \mid \phi, k, \beta) 
    = \frac{1 + (1-q)\cos(k\phi + \beta - m\pi)}{2}.
\end{equation*}

In the following, we describe the process of a single Bayesian update for converting a prior distribution $p(\phi)$ into a posterior distribution $p(\phi \mid m) \propto p(\phi) p(m \mid \phi, k, \beta) $ on receipt of a measurement $m$. For brevity we describe only a single update, however the single update generalises trivially by iteratively setting the prior to the posterior from the previous update, as described in Section~\ref{sec:BQPE}.

\subsubsection*{Fourier Distribution}
Suppose the prior is represented by a Fourier distribution
\begin{equation*}
    p(\phi) = \frac{1}{2\pi} + \sum_{j=1}^J c_j \cos(j\phi) + s_j \sin(j\phi),
\end{equation*}
where $\{c_j, s_j\}_{j=1}^J$ are the distribution coefficients and $c_j = s_j = 0 \; \forall j$ recovers the uniform distribution (a natural initial prior).

As noted in \cite{OBrien2019} $p(\phi)$ is \textit{conjugate} for the likelihood $p(m \mid \phi, k, \beta)$ meaning the posterior takes the same Fourier form (with $J \to J + k$) and can be calculated analytically using the standard trigonometric product-sum formulae.

We note that the circular moments of a Fourier distribution are also tractable
\begin{align}
    \mathcal{M}_j := \mathbb{E}_{p(\phi)}[e^{ij\phi}] &= \mathbb{E}_{p(\phi)}[\cos(j\phi) + i\sin(j\phi)], \nonumber \\
    &= \pi(c_j + is_j). \label{eq:fourier_mj}
\end{align}

\subsubsection*{von Mises}

In \cite{vandenBerg2021}, an adaptive procedure was developed where the Fourier posterior was converted to a wrapped Gaussian distribution when the number of coefficients became prohibitively large. In this work, we advocate for the use of a von Mises distribution over the wrapped Gaussian as the von Mises distribution is the maximum entropy distribution for a wrapped random variable (with specified $\mathcal{M}_1$) and has a tractable density function
\begin{equation*}
    p(\phi) = \frac{\exp[\kappa \cos(\phi - \mu)]}{2\pi I_0(\kappa)},
\end{equation*}
where $\mu$ is the mean, $\kappa$ is a precision parameter and $I_j$ is a modified Bessel function of the first kind and order $j \in \mathbb{N}$. Note that for sharply peaked distributions we have a Gaussian approximation $p(\phi) \approx \text{N}(\phi \mid \mu, \sqrt{1/\kappa})$.

As with Fourier distributions, the circular moments are tractable
\begin{align}\label{eq:vm_mj}
    \mathcal{M}_j = \frac{I_{|j|}(\kappa)}{I_{0}(\kappa)}e^{ij\mu} = \frac{I_{|j|}(\kappa)}{I_{0}(\kappa)}(\cos(j\mu) + i\sin(j\mu)).
\end{align}
This identity can be used to find the first circular moment $\mathcal{M}_1$ of the posterior $p(\phi \mid m) \propto p(\phi)p(m \mid \phi, k, \beta)$. The von Mises distribution is not conjugate for $p(m \mid \phi, k, \beta)$ and therefore the true posterior is not von Mises. However, we can do an approximate moment-matching update (as in \cite{Wiebe2016, vandenBerg2021}) where we find an approximate posterior with unique parameters $\mu, \kappa$ that give a von Mises distribution with first circular moment $\mathcal{M}_1$ matching that of the true posterior $p(\phi \mid m)$. This procedure resorts to inverting \eqref{eq:vm_mj} which can be done efficiently numerically.

We can use the same moment-matching technique to convert a Fourier distribution into a von Mises distribution (with matching $\mathcal{M}_1$). All implementation details can be found in \texttt{phayes} \cite{phayes}.

\subsection{Parameter selection}
\label{app:PS}

As described in \ref{subsec:PS}, we consider the specific (but well-motivated) choice of negative circular variance utility function
\begin{equation*}
    U_C(k, \beta) = \sum_{m \in \{0,1\}} -\text{Var}_C[\phi \mid m, k, \beta] p(m \mid k, \beta).
\end{equation*}
where $\text{Var}_{C}[\phi \mid m, k, \beta] =  1 - |\mathbb{E}_{p(\phi \mid m, k, \beta)}[e^{i\phi}]| \in [0,1]$. The goal is to find $k^*, \beta^* = \argmax_{k, \beta} U_C(k, \beta)$.

It can be shown using (\ref{eq:fourier_mj}-\ref{eq:vm_mj}) that for both a Fourier or von Mises prior, the posterior first circular moment takes the form
\begin{align*}
    \mathbb{E}_{p(\phi \mid m, k, \beta)}[e^{i\phi}] = \frac{
    \bar{a} + \bar{b}\cos(\beta - m \pi) + \bar{c}\sin(\beta - m\pi)
    }{p(m \mid k, \beta)}
    + i
    \frac{\bar{d} + \bar{e}\cos(\beta - m \pi) + \bar{f}\sin(\beta - m\pi))
    }{p(m \mid k, \beta)},
\end{align*}
where $\bar{a}, \dots, \bar{f}$ are real values independent of $\beta$ but dependent on $k$.

Thus the circular variance utility for a given value of $k$ becomes
\begin{equation*}
    U_C(\beta \mid k) = -1 + \sqrt{f^+(\beta)} + \sqrt{f^-(\beta)},
\end{equation*}
where $f^\pm(\beta) = a \pm b\cos(\beta) \pm c\sin(\beta) + d\cos(2\beta) + e\sin(2\beta)$ for new real-valued, $k$ dependent coefficients $a, \dots, e$.

This equation can be maximised in $\beta$ analytically by considering $\partial_\beta U_C(\beta \mid k) = 0$. After multiplying by $\sqrt{f^+(\beta) f^-(\beta)}$, squaring both sides (twice) and simplifying terms, the result is a cubic polynomial in $\cos^2(\beta)$ which can be solved easily. The roots can then be checked to find the maximiser. A \texttt{sympy} \cite{sympy} document listing these equations explicitly can be found in \texttt{phayes} \cite{phayes}.

The aforementioned procedure maximises $U_C(\beta \mid k)$, i.e. for a given $k$. For a Fourier prior this process can simply be repeated for each $k \in \{1, \dots, J\}$ maintaining the same $O(J)$ complexity of a single posterior update. In the von Mises case, we find that checking a small range of $k$ on either side of $\text{Var}_H[\phi]^{-\frac12}$ finds the maximiser.

\subsection{Example posteriors} 
\label{app:posteriors}

\begin{figure}
    \centering
    \includegraphics[width=0.9\hsize]{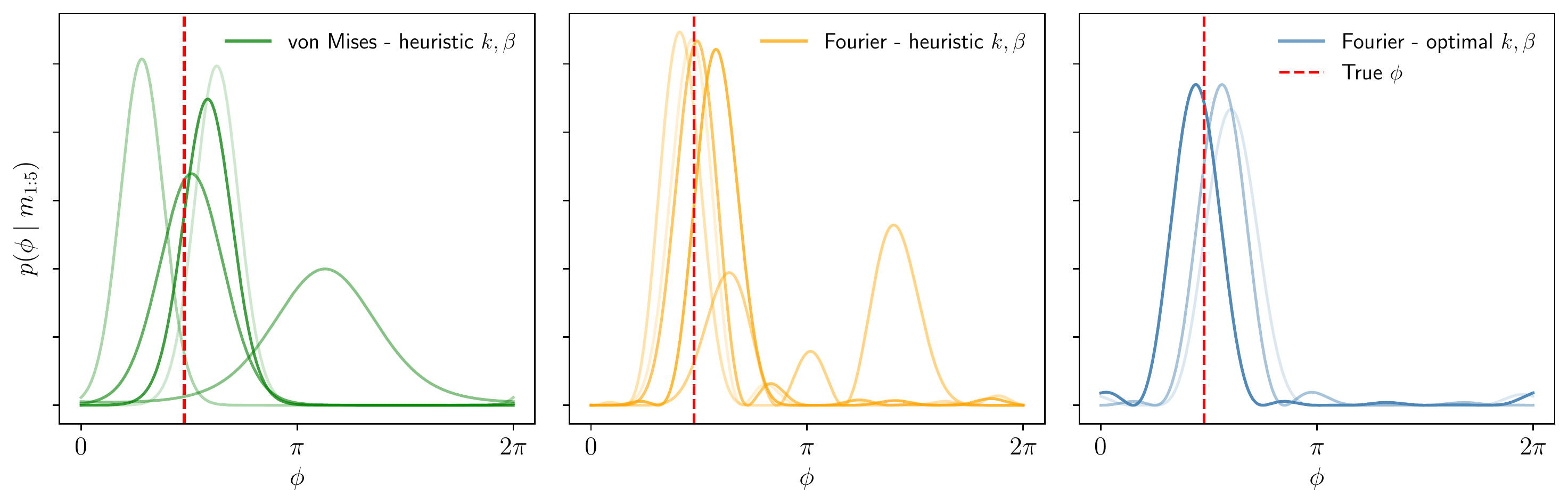}
    \caption{
        Example posteriors for the three Bayesian phase estimation approaches discussed in \ref{subsec:PS} and Figure~\ref{fig:bpe_param_selection}. The posteriors are from noiseless synthetic phases estimation experiments and are displayed after five experiments have been measured. Each curve represents one of five independent runs with a different pseudo-random seed. Note that with just five measurements the adaptive approach never converts to von Mises.
    }
    \label{fig:posterior5}
\end{figure}

In Fig.~\ref{fig:posterior5}, we display some example posteriors for a small synthetic phase experiment where five measurements are collected. Observe that the optimal parameter selection allows the posterior to quickly hone in on the correct phase, whilst the heuristic parameter with the analytical Fourier posterior struggles to rule out incorrect modes. The von Mises approach (and similarly Gaussian-based approaches) are uni-modal by design and therefore have to retain a large variance. We also observe over the repeated experiments (and in Fig.~\ref{fig:bpe_param_selection}) that the uni-modal assumption can result in the correct phase being missed entirely. The optimal parameter selection's ability to eliminate incorrect modes means that we more robustly converge to the correct phase and also that the posterior is better approximated by a von Mises distribution when the adaptive procedure reaches conversion time.

\subsection{Noiseless simulation of Bayesian QPE}
\label{app:noiseless_sim}

\begin{figure}
    \centering
    \includegraphics[width=0.90\hsize]{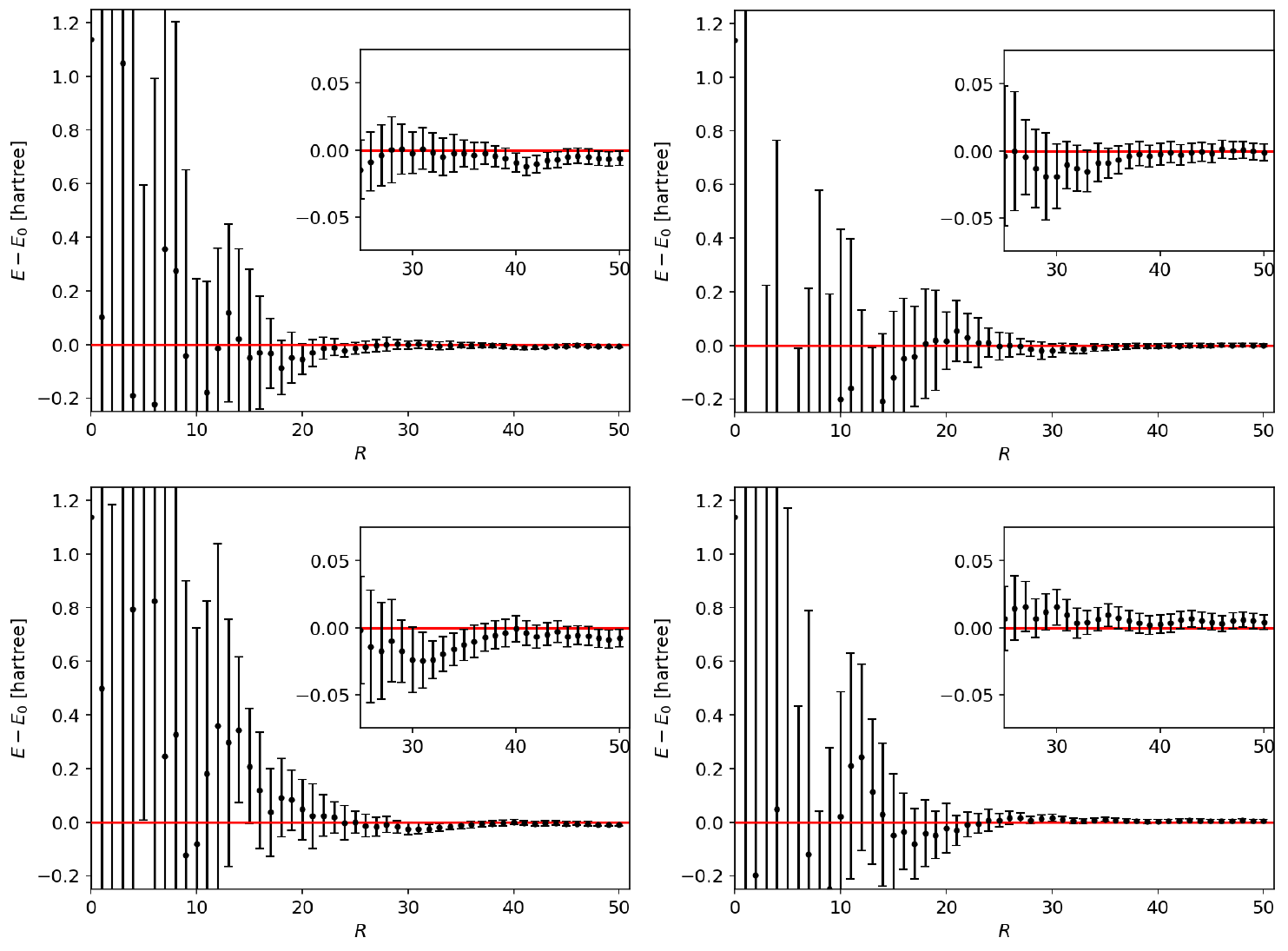}
    \caption{
        Energy estimate plotted along the number of experiments calculated with a noiseless quantum simulator. The four panels correspond to different runs with the same computational setup as those given in Sec. \ref{subsec:bayesian_qpe}.
    }
    \label{fig:noiseless_results}
\end{figure}

To show how much energy estimates vary among different instances of Bayesian QPE experiments, we exemplify the results from the noiseless quantum simulator.
The computational setup is the same as that of the Bayesian QPE given in Sec. \ref{subsec:bayesian_qpe}.
The estimated energies $E$ from the Bayesian QPE simulations agree with the exact ground state energy $E_{0}$ within $\pm \sqrt{\mathrm{Var}_{H}[E]}$ after 50 Bayesian updates as shown in Fig.~\ref{fig:noiseless_results}.
Such a reliable convergence is due to the optimal parameter selection method.
The results also confirm that the use of Hartree-Fock state as an input does not lead to significant systematic error in estimating the ground state energy with the precision we require.

\section{Encoding and compiling the QPE circuit with $\llbracket6,4,2\rrbracket$ code}
\label{app:QED}
\setcounter{figure}{0}

Recall that the unencoded QPE circuit~\eqref{circ:hadamard} is given by
\begin{align}
\label{circ:QPE}
\begin{array}{c}
\Qcircuit @C=.3em @R=1.em {
    \lstick{1: \ket{+}}
    & \qw
    & \qw
    & \ctrl{1}
    & \ctrl{1}
    & \gate{R_{Z}(\beta)}
    & \measureD{X}
    \\
    \lstick{2: \ket{0}}
    & \qw
    & \qw
    & \multigate{1}{u^{ks}}
    & \multigate{1}{v}
    & \qw
    & \qw
    \\
    \lstick{3: \ket{0}}
    & \qw
    & \qw
    & \ghost{u^{ks}}
    & \ghost{v}
    & \qw
    & \qw
    \\
    \lstick{4: \ket{+}}
    & \qw
    & \qw
    & \qw
    & \qw
    & \qw
    & \qw
}
\end{array}
\end{align}
Note the fourth qubit is absent in the experiments with unencoded circuits, but it is added so that the circuit is encoded with $\llbracket n+2,n,2\rrbracket$ code, where $n$ has to be even. The reason why it is initialized to $\ket{+}$ will become clear later [Eq.~\eqref{eq:encode_ctrlU}].
The control unitary operations are
\begin{align}
\label{eq:ctrlU}
    \begin{array}{c}
        \Qcircuit @C=.3em @R=1.5em {
        & \ctrl{1}
        & \qw
        \\
        &\multigate{1}{u}
        & \qw
        \\
        &\ghost{u}
        & \qw
        }
    \end{array}
    ~
    =
    ~
    ~
    ~
    \begin{array}{c}
        \Qcircuit @C=.3em @R=0.8em {
        & \ctrl{1}
        & \ctrl{2}
        & \ctrl{1}
        & \qw
        \\
        &\gate{R_{Z}(\tfrac{2h_{1}t}{s})}
        & \qw
        & \multigate{1}{R_{YY}(\tfrac{2h_{3}t}{s})}
        & \qw
        \\
        & \qw
        &\gate{R_{Z}(\tfrac{2h_{2}t}{s})}
        & \ghost{R_{YY}(\tfrac{2h_{3}t}{s})}
        & \qw
        }
    \end{array}
    =
    \hspace{2em}
    \begin{array}{c}
        \Qcircuit @C=.3em @R=1.2em {
        & \qw
        & \multigate{1}{\begin{array}{c}Z\\ Z\end{array}}
        & \multigate{2}{\begin{array}{c}Z\\ \vspace{5mm} \\ Z\end{array}}
        & \qw
        & \multigate{2}{\begin{array}{c}Z\\ \vspace{-1mm} \\ Y\\ \vspace{-1mm} \\ Y\end{array}}
        & \qw
        \\
        & \gate{Z}
        & \ghost{\begin{array}{c}Z\\ Z\end{array}}
        & \ghost{\begin{array}{c}Z\\ \vspace{5mm} \\ Z\end{array}}
        & \multigate{1}{\begin{array}{c}Y\\ Y\end{array}}
        & \ghost{\begin{array}{c}Z\\ Y \\ Y\end{array}}
        & \qw
        \\
        & \gate{Z}
        & \qw
        & \ghost{\begin{array}{c}Z\\ \vspace{5mm} \\ Z\end{array}}
        & \ghost{\begin{array}{c}Y\\ Y\end{array}}
        & \ghost{\begin{array}{c}Z\\ Y\\ Y\end{array}}
        & \qw
        }
    \end{array}
\end{align}
and
\begin{align}
\label{eq:ctrlV}
    \begin{array}{c}
        \Qcircuit @C=.3em @R=1.3em {
        & \ctrl{1}
        & \qw
        \\
        &\multigate{1}{v}
        & \qw
        \\
        &\ghost{v}
        & \qw
        }
    \end{array}
    ~
    =
    ~
    ~
    ~
    \begin{array}{c}
        \Qcircuit @C=.3em @R=1.0em {
        & \ctrl{1}
        & \gate{R_{Z}(-h_{4}kt)}
        & \qw
        \\
        & \multigate{1}{R_{ZZ}(2h_{2}kt)}
        & \qw
        & \qw
        \\
        & \ghost{R_{ZZ}(2h_{2}kt)}
        & \qw
        & \qw
        }
    \end{array}
    =
    \hspace{2em}
    \begin{array}{c}
        \Qcircuit @C=.3em @R=1.2em {
        & \qw
        & \multigate{2}{\begin{array}{c}Z\\ \vspace{5mm} \\ Z\end{array}}
        & \gate{Z}
        & \qw
        \\
        & \multigate{1}{\begin{array}{c}Z\\ Z\end{array}}
        & \ghost{\begin{array}{c}Z\\ \vspace{-1mm} \\Z\\ \vspace{-1mm} \\Z\end{array}}
        & \qw
        & \qw
        \\
        & \ghost{\begin{array}{c}Z\\ Z\end{array}}
        & \ghost{\begin{array}{c}Z\\ \vspace{5mm} \\ Z\end{array}}
        & \qw
        & \qw
        }
    \end{array}
\end{align}
In the second equality of each equation, we note that a controlled Pauli exponential operator ctrl-$R_P(\theta)$ can be decomposed into a product of Pauli exponential operators using the following identity,
\begin{align}
    \text{ctrl$_{i}$-}R_{P_j}(\theta) = R_{P_j}(\theta/2)R_{Z_iP_j}(-\theta/2).
\end{align}
Also, we introduced the shorthand notation of a ctrl-$R_P(\theta)$ gate that shows only the Pauli operator $P$ on the exponent with the angle $\theta$ suppressed.
Note that $R_Z(\beta)$ in~\eqref{circ:QPE} can be absorbed into the ctrl-$v$~\eqref{eq:ctrlV}.

\vspace{1em}

We discuss each component of encoded circuit~\eqref{fig:logical_circ} and how they are compiled to the universal gates~\eqref{eq:universal_gates}.
For the convenience of discussion, we reproduce the encoded circuit here.
\begin{align}
\nonumber
    \begin{array}{c}
        \Qcircuit @C=.5em @R=0.7em {
            \\
            \\
            \lstick{1}
            & \multigate{7}{\rotatebox[origin=l]{90}{State preparation}}
            & \qw
            \\
            \lstick{2}
            & \ghost{\rotatebox[origin=l]{90}{State preparation}}
            & \qw
            \\
            \lstick{3}
            & \ghost{\rotatebox[origin=l]{90}{State preparation}}
            & \qw
            \\
            \lstick{4}
            & \ghost{\rotatebox[origin=l]{90}{State preparation}}
            & \qw
            \\
            \lstick{a_X}
            & \ghost{\rotatebox[origin=l]{90}{State preparation}}
            & \qw
            \\
            \lstick{a_Z}
            & \ghost{\rotatebox[origin=l]{90}{State preparation}}
            & \qw
            \\
            \lstick{\ket{0}}
            & \ghost{\rotatebox[origin=l]{90}{State preparation}}
            & \cw
            \\
            \lstick{\ket{0}}
            & \ghost{\rotatebox[origin=l]{90}{State preparation}}
            & \cw
        }
    \end{array}
    \left[
    \left[
    \begin{array}{c}
        \Qcircuit @C=.5em @R=.7em {
            & \multigate{5}{\rotatebox[origin=l]{90}{$\overline{\text{ctrl-}u}$}}
            & \qw
            \\
            & \ghost{\rotatebox[origin=l]{90}{$\overline{\text{ctrl-}u}$}}
            & \qw
            \\
            & \ghost{\rotatebox[origin=l]{90}{$\overline{\text{ctrl-}u}$}}
            & \qw
            \\
            & \ghost{\rotatebox[origin=l]{90}{$\overline{\text{ctrl-}u}$}}
            & \qw
            \\
            & \ghost{\rotatebox[origin=l]{90}{$\overline{\text{ctrl-}u}$}}
            & \qw
            \\
            & \ghost{\rotatebox[origin=l]{90}{$\overline{\text{ctrl-}u}$}}
            & \qw
            \\
            \\
            \\
        }
    \end{array}
    \right]^{\frac{f}{2}}
    \begin{array}{c}
        \Qcircuit @C=.5em @R=0.7em {
            & \multigate{5}{S_X}
            & \qw
            \\
            & \ghost{S_X}
            & \qw
            \\
            & \ghost{S_X}
            & \qw
            \\
            & \ghost{S_X}
            & \qw
            \\
            & \ghost{S_X}
            & \qw
            \\
            & \ghost{S_X}
            & \qw
            \\
            \\
            \\
        }
    \end{array}
    \left[
    \begin{array}{c}
        \Qcircuit @C=.5em @R=.7em {
            & \multigate{5}{\rotatebox[origin=l]{90}{$\overline{\text{ctrl-}u}$}}
            & \qw
            \\
            & \ghost{\rotatebox[origin=l]{90}{$\overline{\text{ctrl-}u}$}}
            & \qw
            \\
            & \ghost{\rotatebox[origin=l]{90}{$\overline{\text{ctrl-}u}$}}
            & \qw
            \\
            & \ghost{\rotatebox[origin=l]{90}{$\overline{\text{ctrl-}u}$}}
            & \qw
            \\
            & \ghost{\rotatebox[origin=l]{90}{$\overline{\text{ctrl-}u}$}}
            & \qw
            \\
            & \ghost{\rotatebox[origin=l]{90}{$\overline{\text{ctrl-}u}$}}
            & \qw
            \\
            \\
            \\
        }
    \end{array}
    \right]^{\frac{f}{2}}
    \begin{array}{c}
        \Qcircuit @C=.5em @R=.7em {
            \\
            \\
            &
            & \multigate{7}{\rotatebox[origin=l]{90}{Syndrome meas.}}
            & \qw
            \\
            &
            & \ghost{\rotatebox[origin=l]{90}{Syndrome meas.}}
            & \qw
            \\
            &
            & \ghost{\rotatebox[origin=l]{90}{Syndrome meas.}}
            & \qw
            \\
            &
            & \ghost{\rotatebox[origin=l]{90}{Syndrome meas.}}
            & \qw
            \\
            &
            & \ghost{\rotatebox[origin=l]{90}{Syndrome meas.}}
            & \qw
            \\
            &
            & \ghost{\rotatebox[origin=l]{90}{Syndrome meas.}}
            & \qw
            \\
            & \lstick{\ket{0}}
            & \ghost{\rotatebox[origin=l]{90}{Syndrome meas.}}
            & \cw
            \\
            & \lstick{\ket{0}}
            & \ghost{\rotatebox[origin=l]{90}{Syndrome meas.}}
            & \cw
        }
    \end{array}
    \right]^{\frac{ks}{f}}
    \begin{array}{c}
    \vspace{-2em}
        \Qcircuit @C=.5em @R=0.7em {
            & \multigate{5}{\rotatebox[origin=l]{90}{$\overline{\text{ctrl-}v}$}}
            & \multigate{7}{\rotatebox[origin=l]{90}{Measurement}}
            & \cw
            \\
            & \ghost{\rotatebox[origin=l]{90}{$\overline{\text{ctrl-}v}$}}
            & \ghost{\rotatebox[origin=l]{90}{Measurement}}
            & \cw
            \\
            & \ghost{\rotatebox[origin=l]{90}{$\overline{\text{ctrl-}v}$}}
            & \ghost{\rotatebox[origin=l]{90}{Measurement}}
            & \cw
            \\
            & \ghost{\rotatebox[origin=l]{90}{$\overline{\text{ctrl-}v}$}}
            & \ghost{\rotatebox[origin=l]{90}{Measurement}}
            & \cw
            \\
            & \ghost{\rotatebox[origin=l]{90}{$\overline{\text{ctrl-}v}$}}
            & \ghost{\rotatebox[origin=l]{90}{Measurement}}
            & \cw
            \\
            & \ghost{\rotatebox[origin=l]{90}{$\overline{\text{ctrl-}v}$}}
            & \ghost{\rotatebox[origin=l]{90}{Measurement}}
            & \cw
            \\
            &\lstick{\ket{0}}
            & \ghost{\rotatebox[origin=l]{90}{Measurement}}
            & \cw
            \\
            &\lstick{\ket{0}}
            & \ghost{\rotatebox[origin=l]{90}{Measurement}}
            & \cw
        }
    \end{array}
\end{align}

\subsection{State preparation, syndrome measurements, and final measurement}

Fault-tolerant protocols are known for state preparation, syndrome measurements, and final measurement in $\llbracket n+2,n,2\rrbracket$ QED code. Here, we briefly comment on each primitive by referring to the preceding literature for further details.

\begin{itemize}
\item The Hartree-Fock state $\ket{+00+}$ (exact ground state $e^{-\im \alpha Y_{2}X_{3}/2}\ket{+00+}$ with $\alpha = -0.07113$) is fault tolerantly encoded following Appendix D of~\cite{Chao2018}. It uses 9 two-qubit ($R_{ZZ}$) gates and 2 ancillary qubits to detect faults. 

\item The fault-tolerant syndrome measurements of $S_X$ and $S_Z$ are performed in the form proposed in~\cite{Self2022}, which uses 12 two-qubit gates and 2 ancillary qubits.

\item The final measurement is based on the implementation in~\cite{Chao2018}. The primitive consists of syndrome measurement of $S_Z$ stabilizer and destructive $X$ measurements on all the physical qubits. If no errors are detected, the measurement outcomes are post-processed to extract the observable. There are 8 two-qubit gates and 2 ancillary qubits in the measurement circuit.

\end{itemize}

\subsection{Logical unitary operations}

Converting ctrl-$v$~\eqref{eq:ctrlV} to $\overline{\text{ctrl-}v}$ and compile it to logical gates is straightforward.
We focus on how to compile $\overline{\text{ctrl-}u}$.
We first pull out $I\otimes S\otimes S\otimes I$ and $I\otimes S^\dag\otimes S^\dag\otimes I$ in \eqref{eq:ctrlU} to convert $Y$ operators into $X$ operators,
\begin{align}
\label{eq:ctrlU_+}
    \begin{array}{c}
        \Qcircuit @C=.3em @R=1.5em {
        \lstick{1}
        & \ctrl{1}
        & \qw
        \\
        \lstick{2}
        &\multigate{1}{u}
        & \qw
        \\
        \lstick{3}
        &\ghost{u}
        & \qw
        \\
        &\lstick{4: \ket{+}}
        & \qw
        }
    \end{array}
    ~
    ~
    =
    \hspace{1.5em}
    \begin{array}{c}
        \Qcircuit @C=.3em @R=1.2em {
        & \qw
        & \qw
        & \multigate{1}{\begin{array}{c}Z\\ Z\end{array}}
        & \multigate{2}{\begin{array}{c}Z\\ \vspace{5mm} \\ Z\end{array}}
        & \qw
        & \multigate{2}{\begin{array}{c}Z\\ \vspace{-1mm} \\ X\\ \vspace{-1mm} \\ X\end{array}}
        & \qw
        \\
        & \gate{S^\dag}
        & \gate{Z}
        & \ghost{\begin{array}{c}Z\\ Z\end{array}}
        & \ghost{\begin{array}{c}Z\\ \vspace{5mm} \\ Z\end{array}}
        & \multigate{1}{\begin{array}{c}X\\ X\end{array}}
        & \ghost{\begin{array}{c}Z\\ X \\ X\end{array}}
        & \gate{S}
        & \qw
        \\
        & \gate{S^\dag}
        & \gate{Z}
        & \qw
        & \ghost{\begin{array}{c}Z\\ \vspace{5mm} \\ Z\end{array}}
        & \ghost{\begin{array}{c}X\\ X\end{array}}
        & \ghost{\begin{array}{c}Z\\ X\\ X\end{array}}
        & \gate{S}
        & \qw
        \\
        &\lstick{\ket{+}}
        & \qw
        & \qw
        & \qw
        & \qw
        & \qw
        & \qw
        & \qw
        }
    \end{array}
\end{align}
which simplifies the compilation as we will see momentarily. The operators $I\otimes S\otimes S\otimes I$ and $I\otimes S^\dag\otimes S^\dag\otimes I$ can be absorbed either in the initial state or the final state without affecting the measurement outcome.

Using Eq.~\eqref{eq:logical}, we find the encoded form of~\eqref{eq:ctrlU_+},
\begin{align}
\label{eq:encode_ctrlU}
    \begin{array}{c}
        \Qcircuit @C=.3em @R=.6em {
            \lstick{1}
            & \multigate{5}{\rotatebox[origin=l]{90}{$\overline{\text{ctrl-}u}$}}
            & \qw
            \\
            \lstick{2}
            & \ghost{\rotatebox[origin=l]{90}{$\overline{\text{ctrl-}u}$}}
            & \qw
            \\
            \lstick{3}
            & \ghost{\rotatebox[origin=l]{90}{$\overline{\text{ctrl-}u}$}}
            & \qw
            \\
            \lstick{4}
            & \ghost{\rotatebox[origin=l]{90}{$\overline{\text{ctrl-}u}$}}
            & \qw
            \\
            \lstick{a_X}
            & \ghost{\rotatebox[origin=l]{90}{$\overline{\text{ctrl-}u}$}}
            & \qw
            \\
            \lstick{a_Z}
            & \ghost{\rotatebox[origin=l]{90}{$\overline{\text{ctrl-}u}$}}
            & \qw
        }
    \end{array}
    ~
    =
    \hspace{2em}
    \begin{array}{c}
        \Qcircuit @C=.3em @R=0.6em {
        & \qw
        & \multigate{1}{\begin{array}{c}Z \\ Z \end{array}}
        & \multigate{2}{\begin{array}{c}Z\\ \vspace{0.5mm} \\Z \end{array}}
        & \qw
        & \multigate{5}{\begin{array}{c}Y \\ \vspace{17mm} \\ Y \end{array}}
        & \qw
        \\
        & \multigate{4}{\begin{array}{c}Z\\ \vspace{12mm} \\Z \end{array}}
        & \ghost{\begin{array}{c}Z\\ Z \end{array}}
        & \ghost{\begin{array}{c}Z\\ \vspace{0.5mm} \\Z \end{array}}
        & \multigate{1}{\begin{array}{c}X\\ X \end{array}}
        & \ghost{\begin{array}{c}Y \\ \vspace{17mm} \\ Y \end{array}}
        & \qw
        \\
        & \ghost{\begin{array}{c}Z\\ \vspace{12mm} \\Z \end{array}}
        & \multigate{3}{\begin{array}{c}Z\\ \vspace{6mm} \\Z \end{array}}
        & \ghost{\begin{array}{c}Z\\ \vspace{0.5mm} \\Z \end{array}}
        & \ghost{\begin{array}{c}X\\ X \end{array}}
        & \ghost{\begin{array}{c}Y \\ \vspace{17mm} \\ Y \end{array}}
        & \qw
        \\
        & \ghost{\begin{array}{c}Z\\ \vspace{12mm} \\Z \end{array}}
        & \ghost{\begin{array}{c}Z\\ \vspace{0.5mm} \\Z \end{array}}
        & \qw
        & \qw
        & \ghost{\begin{array}{c}Y \\ \\ \\ \vspace{1.5mm} \\ Y \end{array}}
        & \qw
        \\
        & \ghost{\begin{array}{c}Z\\ \vspace{12mm} \\Z \end{array}}
        & \ghost{\begin{array}{c}Z\\ \vspace{0.5mm} \\Z \end{array}}
        & \qw
        & \qw
        & \ghost{\begin{array}{c}Y \\ \\ \\ \vspace{1.5mm} \\ Y \end{array}}
        & \qw
        \\
        & \ghost{\begin{array}{c}Z\\ \vspace{12mm} \\Z \end{array}}
        & \ghost{\begin{array}{c}Z\\ \vspace{0.5mm} \\Z \end{array}}
        & \qw
        & \qw
        & \ghost{\begin{array}{c}Y \\ \\ \\ \vspace{1.5mm} \\ Y \end{array}}
        & \qw
        }
    \end{array}
\end{align}
where the stabilizer condition $S_X=S_Z=1$ is used to reduce the weight of Pauli operators. Furthermore, we exploit the fourth logical qubit initialize to $\ket{\overline{+}}$, which implies $\overline{X}_4=1$. Then, we find the reduction of the operator,
\begin{equation}
    \overline{Z}_1 \overline{X}_2 \overline{X}_3
    = Z_1 Z_{a_Z} X_2 X_3
    = - Y_1 Y_{a_Z}S_Z\overline{X}_4,
\end{equation}
which is $Y_1 Y_{a_Z}$ in the subspace such that $S_X=S_Z=\overline{X}_4=1$ holds.
Thus, we have compiled all the logical operators to two-qubit Pauli exponential operators to obtain~\eqref{eq:encode_ctrlU}, which are readily implemented by the native two-qubit gate, $R_{ZZ}(\theta)$ and single-qubit gates. Each $\overline{\text{ctrl-}u}$~\eqref{eq:encode_ctrlU} uses 6 two-qubit gates.

\subsection{Memory error suppression with $X$-stabilizer insertions}
\label{app:memory}

\begin{figure}
    \centering
    \includegraphics[width=0.95\hsize]{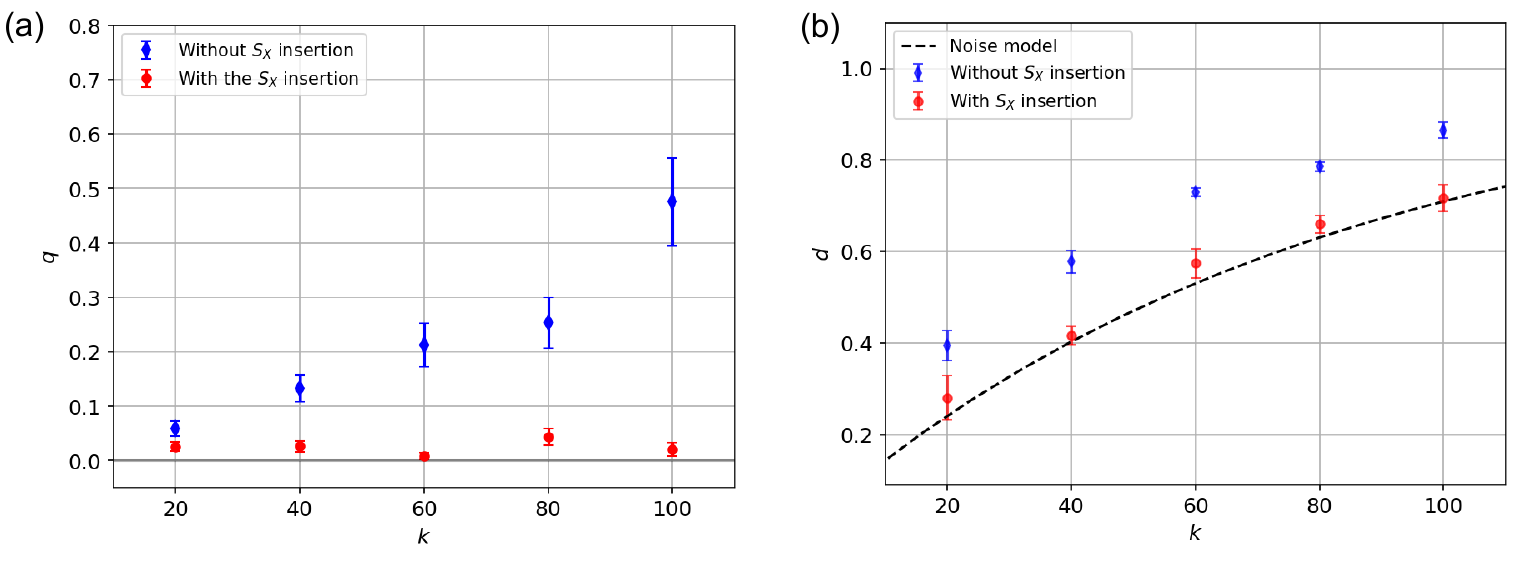}
    \caption{ \label{fig:error_discard_Sx}
        (a) Parameter $q$ of \eqref{eq:fit_func} and (b) discard rate $d$ are shown against $k$ that are evaluated with Quantinuum's emulator of H1--1 quantum computer. The blue and red data points with sampling errors are obtained from the encoded QPE circuit without and with the $X$-stabilizer insertions, respectively. The black dashed curve is drawn using the noise model~\eqref{eq:model_p2_d} with $p_2=1.6\times 10^{-3}$.
    }
\end{figure}

While syndrome measurements can detect any single-qubit errors, if the number of faults in each block between two consecutive measurements is not sufficiently low, they can damage the logical circuit as logical errors. The physical errors can quickly spread across many qubits because the logical operations are not fault-tolerantly implemented.
We found that the physical coherent error has sizable impacts on both the logical error rate and discard rate, as shown in Fig.~\ref{fig:error_discard_Sx}.

The coherent error on the H1 quantum device is mainly modelled by the memory error $e^{i \gamma Z_j}$ with $\gamma\in\mathbb{R}$ and $j\in T$~\cite{H1datasheet}. The errors mostly occur on the qubits during their idling or transportation. As such, the parameter $\gamma$ is roughly characterized by (in the H1 emulator, it is proportional to) the idling or transportation time. Since $X_j$ flips the sign of the exponent in $e^{i \gamma Z_j}$, the memory errors are expected to cancel before and after the inserted $S_X$ in each block.

To alleviate the physical memory error, we insert an $X$-stabilizer, $S_X = \bigotimes_{i\in T}X_i$, in the middle of each block.
While $S_X$ acts as the identity on logical states, we observe that the stabilizer insertion suppresses coherent error, which in turn reduces the discard rate and $q$ of \eqref{eq:fit_func} as shown in Fig.~\ref{fig:error_discard_Sx}.

\section{Suppressing over/under-rotation of $R_{ZZ}$ gates}
\label{app:gate_error_mitigation}
\setcounter{figure}{0}

A small over/under-rotation in $R_{ZZ}(\theta)$ forces us to apply the gate $R_{ZZ}(\theta\pm\delta)$ for some unknown $\delta$. It may not be negligible as the error accumulates similarly to the coherent error.
Since $R_{ZZ}(\pm\delta)$ commutes with the stabilizers $\{S_X,S_Z\}$, the errors are not detected by the syndrome measurements of the $\llbracket6,4,2\rrbracket$ code.

To deal with the misalignment of the angle in $R_{ZZ}(\theta)$, we assume that the imperfect rotation angle is characterized by
\begin{align}
\label{eq:rotation_delta}
    \delta =  \bar{\delta} \frac{\theta}{|\theta|}.
\end{align}
i.e., the misalignment $\delta$ is constant in magnitude and only sensitive to the sign of $\theta$.
To eliminate $\delta$, we modify the time period $t$ of each discretized time step in $U(t)$ to $t_{1}$ and $t_{2}$ for even and odd $k$, respectively. 
For instance, for two consecutive time evolution operators, we make the following adjustment,
\begin{equation}
    \label{eq:u_with_t1_t2}
    \left(
        e^{-\im Ht}
    \right)^{2}
    \to
    e^{-\im Ht_{1}}
    e^{-\im Ht_{2}}
    ,
\end{equation}
with some $t_{1}$ and $t_{2}$ such that $t_{1} + t_{2} = 2t$ and $t_{1}t_{2} < 0$.
It indeed removes the rotation error with~\eqref{eq:rotation_delta} at the cost of larger error from the Lie-Trotter's formula because the modified evolution time obeys $\max(t_1,t_2)> t$.

We tested the protocol on Quantinuum's H1--1 emulator and H1--1 hardware with different sets of $(t_1,t_2)$ summarized in Table~\ref{tab:my_label}.
We see the clear reduction of the phase shift $\omega$ with Eq.~\eqref{eq:u_with_t1_t2}.
In the present QPE experiments, $t_{1} = -0.05\pi$ and $t_{2} = 0.25\pi$ in atomic units are employed.

\begin{table}[tb]
    \centering
    \caption{\label{tab:my_label}
    The phase shift $\omega$ in estimating phase calculated on the emulator (H1-1E) and hardware (H1-1) with different sets of time periods $t_{1}$ and $t_{2}$. $k=60$. The bottom two rows show the results with the protocol~\eqref{eq:u_with_t1_t2}.}
    \begin{tabular}{c|crrr}
        \hline
        & Backend & $t_{1}/\pi$ & $t_{2}/\pi$ & $\omega / 10^{-3}\pi$ \\
        \hline\hline
        1 & H1-1E & $0.30$ & $0.30$ & $0.0\pm0.1$ \\
        \hline
        2 & H1-1 & $0.30$ & $0.30$ & $4.7\pm0.2$ \\
        3 & H1-1 & $0.20$ & $0.20$ & $4.3\pm0.2$ \\
        4 & H1-1 & $0.10$ & $0.10$ & $3.7\pm0.1$ \\
        \hline
        5 & H1-1 & $-0.10$ & $0.50$ & $0.2\pm0.3$ \\
        6 & H1-1 & $-0.05$ & $0.25$ & $0.5\pm0.2$ \\
        \hline
    \end{tabular}
\end{table}

\begin{figure}
    \centering
    \includegraphics[width=0.47\hsize]{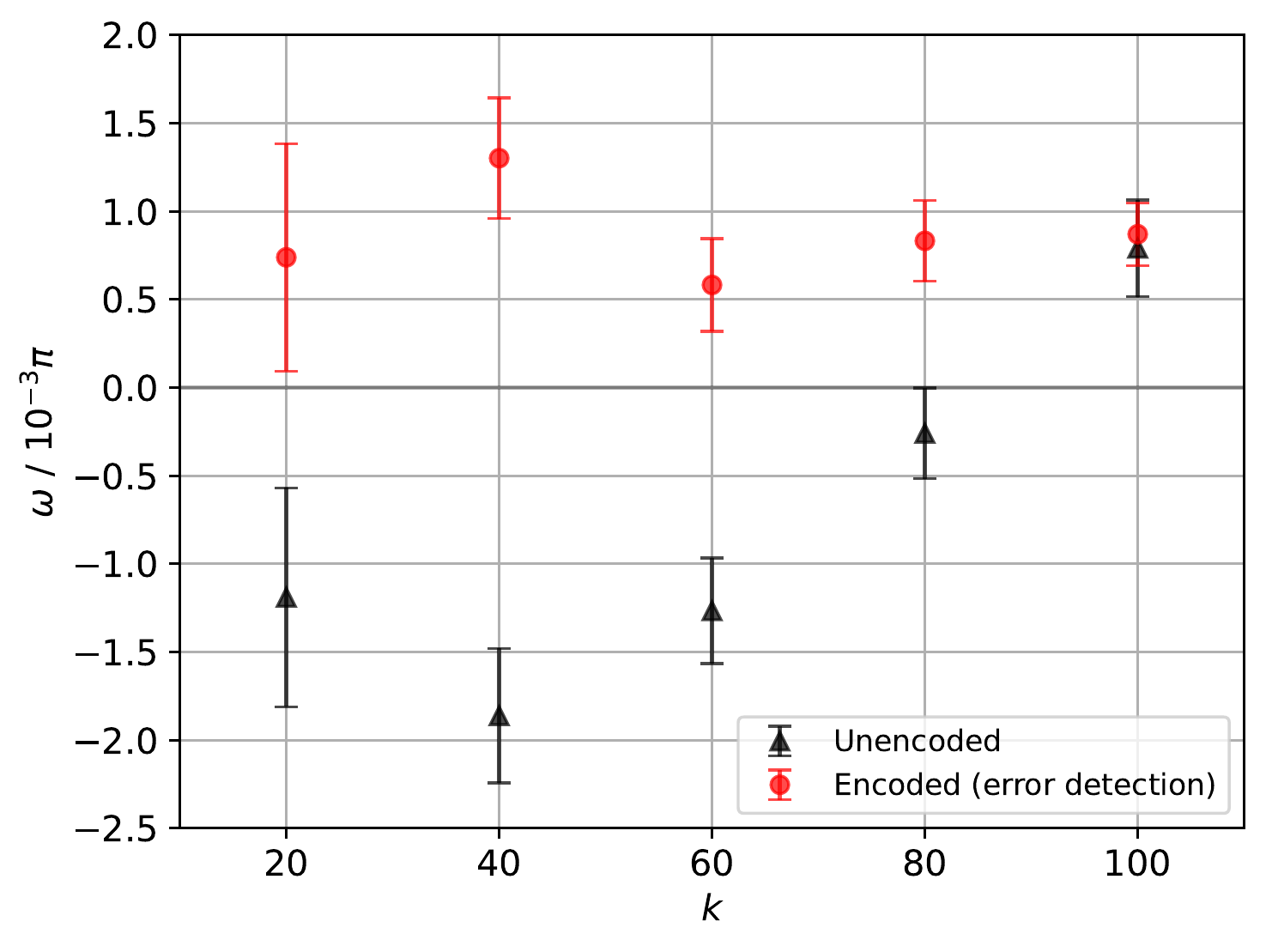}
    \caption{Phase shift $\omega$ in \eqref{eq:fit_func} evaluated for unencoded (red circle) and encoded (black triangle) QPE circuits for each sample point of $k$.}
    \label{fig:benchmark_omega}
\end{figure}

Finally, we show the phase shift $\omega$ in the calibration experiment with and without error detection as discussed in Sec~\ref{subsec:benchmark}.
The protocol~\eqref{eq:u_with_t1_t2} is employed for both cases.
The encoded circuit contains the $S_X$ insertion.
In both unencoded and encoded experiments, the magnitudes of $\omega$ are of order $10^{-3}$.

\end{document}